\documentclass[11pt,a4paper]{article}
\usepackage{jcappub}
\usepackage{natbib}

\newcommand{\cR}{\,\mathrm{{\cal R}}}
\newcommand{\cN}{\,\mathrm{{\cal N}}}
\newcommand{\cG}{\,\mathrm{{\cal G}}}

\newcommand{\be}{\,\begin{equation}}
\newcommand{\ee}{\,\end{equation}}
\newcommand{\abs}[1]{\lvert#1\rvert}

\title{Diffusive propagation of cosmic rays from supernova remnants in the Galaxy. I: spectrum and chemical composition}

\author{Pasquale Blasi and Elena Amato}

\affiliation{INAF/Osservatorio Astrofisico di Arcetri, Largo E. Fermi, 5 - 50125 Firenze, ITALY}

\emailAdd{blasi@arcetri.astro.it, amato@arcetri.astro.it}

\abstract{
In this paper we investigate the effect of stochasticity in the spatial and temporal distribution of supernova remnants on the spectrum and chemical composition of cosmic rays observed at Earth. The calculations are carried out for different choices of the diffusion coefficient $D(E)$ experienced by cosmic rays during propagation in the Galaxy. In particular, at high energies we assume that $D(E)\propto E^{\delta}$, with $\delta=1/3$ and $\delta=0.6$ being the reference scenarios. The large scale distribution of supernova remnants in the Galaxy is modeled following the distribution of pulsars, with and without accounting for the spiral structure of the Galaxy. We find that the stochastic fluctuations induced by the spatial and temporal distribution of supernovae, together with the effect of spallation of nuclei, lead to mild but sensible violations of the simple, leaky-box-inspired rule that the spectrum observed at Earth is $N(E)\propto E^{-\alpha}$ with $\alpha=\gamma+\delta$, where $\gamma$ is the slope of the cosmic ray injection spectrum at the sources. Spallation of nuclei, even with the small rates appropriate for He, may account for small differences in spectral slopes between different nuclei, possibly providing an explanation for the recent CREAM observations. For $\delta=1/3$ we find that the slope of the proton and helium spectra are $\sim 2.67$ and $\sim 2.6$ respectively (with fluctuations depending on the realization of source distribution) at energies around $\sim 1$ TeV (to be compared with the measured values of $2.66\pm 0.02$ and $2.58\pm 0.02$). For $\delta=0.6$ the hardening of the He spectra is not observed.  
The stochastic effects discussed above cannot be found in ordinary propagation calculations, such as GALPROP, where these effects and the point like nature of the sources are not taken into account. We also comment on the effect of time dependence of the escape of cosmic rays from supernova remnants, and of a possible clustering of the sources in superbubbles. In a second paper we will discuss the implications of these different scenarios for the anisotropy of cosmic rays.} 

\begin{document}
\maketitle

\section{Introduction}

Any investigation of the origin of cosmic rays (CRs) has to address some basic questions: 1) where and how are CRs produced/accelerated? 2) what is their chemical composition at the sources? 3) How do they propagate to us? 4) What is the anisotropy that results from source distribution (in space and time) and from propagation? 

After the pioneering proposal of Baade and Zwicky \cite{Baade:1934p884}, the idea that the bulk of Galactic CRs may be produced in supernova remnants (SNRs) has become increasingly more popular. The main reason for this success is that on energetic grounds these astrophysical sources are the most reliable possibility in that an efficiency of $\sim 10\%$ in the form of accelerated particles seems to be able to provide a qualitatively good description of the CR fluxes observed at Earth. The second major reason for the success has come several decades after the original idea was put forward, and relies on the proposal of diffusive shock acceleration (DSA) \cite{Bell:1978p1342,Bell:1978p1344,Blandford:1987p881} as a conversion mechanism from kinetic energy of the expanding supernova blast wave into accelerated particles. This mechanism is seen at work in shocks in the solar system and is based on relatively simple principles. These two ingredients, together with a bulk of observations, showing likely hints of efficient CR acceleration in SNRs, have determined the upgrading of the SNR idea to the rank of a standard paradigm. Yet, the paradigm has not been proven right beyond doubt so far, as discussed recently in Ref. \cite{blasiICATPP}.

The problems with confirming the paradigm originate primarily from addressing all the four questions reported above self-consistently: on one hand the spectra of CRs accelerated in SNRs, according with test-particle DSA are very close to power laws $N(E)\propto E^{-\gamma}$ with slope $\gamma\approx 2$. On the other hand, the spectra measured at Earth require a diffusion coefficient $D(E)\propto E^{\delta}$ with $\delta\sim 0.7$, so that the equilibrium spectrum is $n(E)\propto N(E)/D(E)\propto E^{-\gamma-\delta}=E^{-2.7}$. However the strong diffusion that follows from this line of thought seems to lead to exceedingly large anisotropy \cite{Ptuskin:2006p620}. Non linear effects in DSA make this problem even more severe in that spectra are predicted to be concave and lead to require an even stronger energy dependence of the diffusion coefficient at high energies, as found in Ref. \cite{Berezhko:2007p1010}, where, however, the problem of anisotropy was not discussed. 

It has recently been proposed that the non-linear theory of DSA (hereafter NLDSA) may lead to somewhat steeper spectra if the velocity of the scattering centers is taken into account in solving the transport equation for CRs at SNR shocks \cite{Caprioli:2010p133,Caprioli:2010p789,Ptuskin:2010p1025}. In this case the spectra of accelerated particles may be less concave, and have an approximate power law shape with slope $\approx 2.1-2.2$, thereby implying $\delta \approx 0.5-0.6$. The steeper slope of the injection spectrum follows from both the finite velocity of the scattering waves and from the convolution over time of the acceleration history of SNRs throughout their evolution in the interstellar medium (ISM). It is noteworthy, however, that this mechanism introduces a very disappointing element in that the result depends rather strongly on poorly understood details such as the wave helicity, as well as the reflection and transmission of waves at the shock surface (in principle the effect considered here could even lead to harder spectra).

The most serious limitation to the identification of SNRs as the sources of Galactic CRs is the lack of a firm detection of gamma rays that can unequivocally be interpreted as due to production and decay of neutral pions: in the few cases in which this association appears to be easier, the SNR is close to a molecular cloud (denser target for pp interactions) and typically of old age. In such circumstances the spectrum of accelerated particles is unlikely to be representative of SNRs at their best in terms of CR production. In these few cases, observations show a rather steep spectrum of accelerated particles, quite unlike the ones predicted by NLDSA. Possible exceptions are RX J1713.7-3946 and Tycho, although for the first remnant other problematic aspects may be found \cite{Morlino:2009p140,Ellison:2010p636}. Moreover, a recent analysis of the data on this remnant by the Fermi collaboration \cite{Abdo:2011p1858} suggests that in fact the spectrum of gamma rays is very flat thereby leading to conclude that the emission is likely of leptonic origin. On the other hand the recent gamma ray detection of the Tycho supernova remnant by the VERITAS Cherenkov telescope \cite{2011ApJ...730L..20A} and by the Fermi telescope \cite{2011arXiv1108.0265G} appear to be best fit by a model in which particle acceleration occurs efficiently \cite{2011arXiv1105.6342M}.

As discussed in \cite{Caprioli:2010p789,Ptuskin:2010p1025}, a qualitatively good fit to the spectra of CRs (including nuclei heavier than hydrogen) can be achieved within the context of NLDSA with finite speed of scattering centers (implying $\delta\approx 0.55$) but some spectral features that have been recently found, the so-called discrepant hardenings \cite{Ahn:2010p624} observed by CREAM, are not explained as yet within the context of the SNR paradigm (for a different, environment related explanation, see \cite{ohiraioka:2011p729}). 

From the point of view of propagation in the Galaxy, the standard calculations carried out with GALPROP \cite{GALPROP2011}, or similar propagation codes, suggest that one can obtain a good global fit also with reacceleration models, namely with diffusion coefficient $D(E)\propto E^{1/3}$ and second order Fermi acceleration: the latter is relevant only at low energies, where it allows one to better reproduce the B/C ratio. This scenario would imply a high energy injection spectrum $N(E)\propto E^{-2.4}$, very challenging for NLDSA in SNRs, but certainly preferable from the point of view of anisotropy (see Paper II \cite{paper2}). 

In the present work we use the Green function formalism to describe the propagation of CRs to Earth if the sources are SNRs. These are modeled as discrete sources in space and time, with a spatial distribution and a rate deduced from observations. The effect of discreteness on the spectrum and chemical composition of CRs observed at Earth is the goal of this first paper. In a second paper \cite{paper2} we will discuss the more delicate issue of anisotropy. We concentrate here on 1) the investigation of the spectra and chemical fluctuations induced by the discrete nature of the sources; 2) the chemical composition around the knee; 3) the spectral differences between protons and heavier elements; 4) the effect of extragalactic CRs on the end of the Galactic CR spectrum.

Previous attempts at taking into account the random nature of the sources on the CR spectrum observed at Earth were presented in \cite{Busching:2005p627,Ptuskin:2006p620}. The first paper mainly focuses on the issue of anisotropy and will be discussed in more detail in the forthcoming Paper II. 

In \cite{Busching:2005p627}, the authors use a series expansion of the propagation equation to show that the stochasticity of source distribution induces large temporal fluctuations in the spectrum of primary CR nuclei. Assuming a diffusion coefficient $D(E)\propto E^{0.6}$, they find fluctuations by 20\% on average, but up to 100\%, which leads them to question the goodness of the B/C ratio as an indicator of CR propagation parameters in the Galaxy. However, a diffusion coefficient scaling as $E^{0.6}$ implies a level of anisotropy well in excess of the observations, as discussed in \cite{Ptuskin:2006p620} and in Paper II. If a weaker energy dependence of the diffusion coefficient is considered, the fluctuations found by \cite{Busching:2005p627} are much reduced, weakening also their conclusion.  

The present paper is organized as follows: in \S~\ref{sec:green} we introduce the relevant Green functions; in \S~\ref{sec:simple1} we derive some basic results for the simple case of a homogeneous distribution of sources in the disc of the Galaxy and discuss the limitations of the diffusion formalism for individual sources in \S~\ref{sec:nearby}. In \S~\ref{sec:simple2} we derive simple estimates of the effect of fluctuations on the spectrum for a homogeneous but discrete distribution of sources in the disc. Our results on spectra and chemical composition for a stochastic distribution of the sources in the disc are presented in \S~\ref{sec:results}), where the spiral structure of the Galaxy is first ignored (\S~\ref{sec:res_cyl}) and then taken into account (\S~\ref{sec:res_spiral}). Additional scenarios are discussed in \S~\ref{sec:more}, while we discuss the transition region between galactic and extragalactic CRs in \S~\ref{sec:extragal}. We conclude in \S~\ref{sec:conclusion}.

\section{Relevant Green functions}\label{sec:green}

The diffusive transport of cosmic rays from a point source located at a position $\vec r_{s} = (x_{s},y_{s},z_{s})$ and injecting a spectrum $N(E)$ at a time $t_{s}$ can be written as:
\be
\frac{\partial n_{k}(E,\vec r,t)}{\partial t}=\nabla\left[D_{k}(E)\nabla n_{k}(E,\vec r,t)\right] - \Gamma_{k}^{sp}(E) n_{k}(E,\vec r,t) + N_{k}(E) \delta(t-t_{s})\delta^{3}(\vec r - \vec r_{s}),
\label{eq:transport}
\ee
where $n_{k}(E,\vec r,t)$ is the density of particles of type $k$ (nuclei) with energy $E$ at the location $\vec r$ and time $t$, $D_{k}(E)$ is the diffusion coefficient assumed to be spatially constant and $\Gamma_{k}^{sp}(E) $ is the rate of spallation of nuclei of type $k$ to lead to lighter nuclei. Each source is assumed to produce nuclei of H ($k=1$), He ($k=2$), CNO ($k=3$), Mg-Al-Si ($k=4$) and Fe ($k=5$). The cutoff energy of the proton component, $E_{max}^{H}$ is taken in such a way as to achieve the best fit to the all-particle CR spectrum. For heavier nuclei the maximum energy is chosen to be $Z$ times larger than for H, where $Z=2$ for He, $Z=7$ for CNO, $Z=13$ for Mg-Al-Si and $Z=26$ for Fe nuclei.  All injection spectra are therefore in the form:
\be
N_{k}(E) \propto E^{-\gamma} \exp \left[ -\left( \frac{E}{E_{max,k}} \right)\right],
\ee
where the proportionality constant for each of the nuclear species is determined {\it a posteriori} in order to fit the normalization of the spectra of nuclei of type $k$ as measured at Earth by CREAM at the energy of 1 TeV \cite{Ahn:2010p624,Ahn:2009p1595}. It is worth recalling that the scaling of the maximum energy with the charge $Z$ of the nucleus relies upon the assumption of full ionization of the nucleus. In fact, this condition is achieved only during the acceleration process and the actual maximum energy of heavy nuclei might be less than $Z$ times the maximum energy of protons, because of the difficulty in achieving photo-ionization of atoms through extraction of electrons in the inner atomic shells \cite{morlino:2011mnras}.

The diffusion region is assumed to be a cylinder of infinite radius and half height $H$. Inside this cylinder the diffusion coefficient is constant. The fact that the radius is assumed to be potentially infinite is not a limitation to the calculation because we will see that the source distributions that will be adopted are concentrated in a region of $4-10$ kpc radius (the Galactic disc) and for practical purposes the size of the diffusion region can be taken to be much larger than the radius of the disc. The escape of CRs from this model Galaxy occurs through the upper and lower boundaries. The escape is modeled by assuming that $n_{k}$ vanishes at $z=\pm H$, and the escape flux through the surfaces $z=\pm H$ is described by $D_{k}(E) \frac{\partial n_{k}}{\partial z}|_{z=\pm H}$. 

In this paper we concentrate our attention on particles with energy above $\sim 1$ TeV, therefore in Eq. \ref{eq:transport} we ignored terms of advection and terms of possible second order reacceleration which are both potentially important at much lower energies. We also neglect the contribution to the flux of nuclei deriving from secondary nuclei produced in spallation events. From the point of view of diffusive propagation the calculation presented here is not too different from GALPROP or similar propagation codes. The only two relevant differences are in the more phenomenological way in which spallation is described and in the fact that here we do not impose the free escape boundary condition at some finite radius along the lateral sides of the cylinder. The latter, as discussed above, is not a relevant limitation although it might lead to some corrections for large values of $H$. The former is not important in the present context since we will limit ourselves to primary nuclei, while it might represent a limitation if we were to describe rare nuclei, such as $B$ and $^{10}Be$ or similar secondary products. 

The spallation rate is defined in terms of the gas density in the diffusion region and of the cross section for spallation: $\Gamma_{k}^{sp,k}(E) = n_{gas} c \sigma_{sp,k}$ (the velocity of nuclei has been assumed to equal the speed of light $c$ since all our calculations will only concern ultra-relativistic nuclei). The cross section for spallation of a nucleus $k$ (mass $A_{k}$) can be written as \cite{Horandel:2007p877}:
\be
\sigma_{sp,k}(E) = \alpha_{sp}(E) A_{k}^{\beta_{sp}(E)}~\rm mb,
\ee
where
\be
\alpha_{sp}(E) = 50.44-7.93 \log(E_{eV})+0.61\left[\log(E_{eV})\right]^{2} 
\ee
and
\be
\beta_{sp}(E) = 0.97-0.022\log(E_{eV})
\ee

where $E_{eV}$ is the energy of the nucleus of type $k$ in units of eV. The gas density is non-trivial to define in a homogeneous model as this is. We assume here that the gas density in the disc of the Galaxy (with half-thickness $h$) is $n_{disc}=3$ cm$^{-3}$ while it vanishes in the halo. In this case the mean density experienced by CRs during propagation is approximately $n_{gas}\approx n_{disc}(h/H)$. 

The free Green function (namely without boundary conditions) of Eq.~\ref{eq:transport} can be written as
\be
{\cal G}_{k}^{free}(\vec r, t; \vec r_{s},t_{s}) = \frac{N_{k}(E)}{\left[ 4 \pi D_{k} \tau \right]^{3/2}} \exp\left[ -\Gamma_{k}^{sp}(E) \tau \right]
\exp\left[ -\frac{(\vec r -\vec r_{s})^{2}}{4 D_{k} \tau}\right],
\label{eq:greenfree}
\ee
where $\tau = t-t_{s}$. The Green function that satisfies the correct boundary condition at $z=\pm H$ can be obtained by using the image charge method and can be written as follows:
$$
{\cal G}_{k}(\vec r, t; \vec r_{s},t_{s}) = \frac{N_{k}(E)}{\left[ 4 \pi D_{k} \tau \right]^{3/2}} \exp\left[ -\Gamma_{k}^{sp}(E) \tau \right]
\exp\left[ -\frac{(x-x_{s})^{2}+(y-y_{s})^{2}}{4 D_{k} \tau}\right] \times
$$
\be
\sum_{n=-\infty}^{+\infty} (-1)^{n} \exp \left[ -\frac{(z-z'_{n})^{2}}{4 D_{k} \tau}\right],
\label{eq:green}
\ee
where $z'_{n}=(-1)^{n} z_{s} + 2 n H$ are the $z$ coordinates of the image sources. It is easy to check that ${\cal G}_{k}(x,y,z=\pm H,t;x_{s},y_{s},z_{s},t_{s})=0$ by simply expanding the sum term. In what follows the Earth will be located at $(x,y,z)\equiv (R_{\odot},0,0)$ with $R_{\odot}=8.5$ kpc the distance of the Sun from the center of the Galaxy. It is however useful to keep Eq.~\ref{eq:green} in its most general form because anisotropies are related to the spatial derivatives of the Green function calculated at the detection location. 

It is important to realize that the Green function formalism outlined here allows us to take into account a completely arbitrary spatial distribution of the sources in the Galaxy, and also an arbitrary temporal evolution of the injection of cosmic rays from each source.  

Two instances related to the specific case of SNRs as sources of CRs may clarify the importance of this point. SNRs produce CRs in a rather complicated and not entirely clear way: while the spectrum of CRs accelerated at the external shock is well defined and can be calculated using the theory of NLDSA, the spectrum of particles escaping a remnant to become CRs is much more troublesome in that it depends on how and when CRs escape from the acceleration region. Most accelerated particles in a SNR shock are advected downstream of the shock and are trapped in the expanding shell for the whole duration of the Sedov-Taylor phase, thereby losing energy adiabatically. These particles become CRs in the ISM only after the SNR ends its evolution and the shock dies out. On the other hand, during the Sedov-Taylor phase and in presence of magnetic field amplification and damping, the maximum energy of accelerated particles decreases with time. At any given time the particles with the instantaneous maximum energy may escape the system from upstream. The spectrum of these particles at any given time $t$ is close to a delta function $\delta(E-E_{max}(t))$, where $E_{max}(t)$ is the maximum energy reached at that time. The convolution over time of these peaked spectra leads to a power law injection spectrum \cite{Caprioli:2009p145} which is not directly related to the spectrum of particles at the shock. The global injected spectrum from an individual SNR is likely to be the result of the superposition of the spectrum of particles escaping the SNR from upstream and of those that escape after the end of the Sedov-Taylor phase. The whole process lasts $\sim (3-30)\times 10^{4}$ years, with the low energy particles escaping at later times than the higher energy ones. During the ejecta dominated phase that precedes the adiabatic phase, the maximum energy of accelerated particles grows and escape is not effective. 

In our calculations we consider two different models of injection: 1) burst injection (all particles from a SNR are injected at the same time with a spectrum $N(E)$); 2) continuous injection in which at any time the injected spectrum is $\propto \delta\left[E-E_{max}(t)\right]$. This second choice should qualitatively reflect the statement that escape of particles from the accelerator takes place in such a way that higher energy particles escape earlier in the evolution, while lower energy particles leave later. 

In the first scenario, the Green function in Eq.~\ref{eq:green} is already the solution that we are seeking. The spectrum injected by an individual SNR can be written in the form
\be
N_{k}(E) = \frac{(\gamma_{k}-2) \eta_{k}\epsilon_{kin}}{E_{0,k}^{2}} \frac{1}{\left[ 1-\left( \frac{E_{max,k}}{E_{0,k}}\right)^{-\gamma_{k}+2}\right]} \left(\frac{E}{E_{0}} \right)^{-\gamma_{k}} \exp\left( - \frac{E}{E_{max,k}}\right),
\label{eq:inj}
\ee
where $\eta_{k}$ is the fraction of the kinetic energy of the blast wave, $\epsilon_{kin}$, that goes into accelerated particles of type $k$. The reference energy $E_{0,k}$ is taken to be $1$ GeV for protons while for heavier nuclei its numerical value is not really important since we simply rescale the injected protons spectrum in order to fit the spectra observed at Earth. In Eq.~\ref{eq:inj}, $E_{max,k}$ is the maximum energy of particles of type $k$. In the expression above and in what follows we always assume that the injection spectrum is steeper than $E^{-2}$, since flatter spectra would result in unreasonable choices of the diffusion coefficient, namely in exceedingly large anisotropy and even in the breaking of the regime of diffusive propagation (see below). 

The second model of injection that we consider is introduced, as we already mentioned, in order to take into account, at least in a qualitative way, the fact that escape of particles from the accelerator is expected to occur in a differential way, with higher energy particles escaping first. As discussed in \cite{Caprioli:2009p145,Caprioli:2010p133}, the escape starts at the beginning of the Sedov-Taylor (ST) phase, while during the ejecta dominated (ED) phase of the SNR evolution the particles are confined in the expanding shell, the maximum energy increases with time and the particles lose energy adiabatically. During the ST phase, in the presence of magnetic field amplification, the maximum energy decreases after reaching its maximum value between the end of the ED phase and the beginning of the ST phase. 

The ST phase starts when the mass swept up by the supernova shock equals the mass of the ejecta, $M_{ej}$:
\be
M_{ej} = \frac{4}{3}\pi \rho R_{ST}^{3} \to R_{ST} = \left(\frac{3 M_{ej}}{4\pi \rho} \right)^{1/3}=6.6\times 10^{18}\left(\frac{M_{ej,\odot}}{n_{1}}\right)^{1/3} cm,
\ee
where $R_{ST}$ is the radius of the shell at the beginning of the ST phase, $M_{ej,\odot}$ is the mass of the ejecta in units of solar masses and $n_{1}$ is the gas density in the ISM in units of 1 particle per cubic cm. The kinetic energy of the explosion is related to the mass of the ejecta as
\be
\epsilon_{kin} = \frac{1}{2} M_{ej} u_{ST}^{2} 
\ee
which leads to
\be
 u_{ST} = 10^{9} \left(\frac{\epsilon_{51}}{M_{ej,\odot}} \right)^{1/2} cm/s,
\label{eq:ust}
\ee
where $u_{ST}$ is the velocity of the expanding shell at the beginning of the ST phase. It is interesting to notice that
\be
\epsilon_{kin} = \frac{1}{2} M_{ej} u_{ST}^{2} = \left(\frac{1}{2}\rho u_{ST}^{3} \right) \left( 4 \pi R_{ST}^{2} \right) T_{ST}\ .
\label{eq:ekin}
\ee
This equation, together with Eq.~\ref{eq:ust}, gives an expression for the time $T_{ST}$ at which the ST phase approximately starts:
\be
T_{ST}=\frac{1}{3}\frac{R_{ST}}{u_{ST}} \approx 70 yr \frac{M_{ej,\odot}^{5/6}}{\epsilon_{51}^{1/2}n_{1}^{1/3}}.
\ee

During the ST phase the radius and velocity of the shell change as
\be
R_{sh}(t) = R_{ST} \left( \frac{t}{T_{S}}\right)^{2/5} ~~~~~~~ u_{sh}(t) = \frac{6}{5} u_{ST} \left( \frac{t}{T_{ST}}\right)^{-3/5}.
\ee

We assume that escape of CRs from the SNR occurs from upstream of the shock, therefore the spectrum of escaping particles at any given time is very much peaked around $E_{max}(t)$ (we omit here the index $k$ that identifies the type of nucleus) and for simplicity we assume $Q(E,t) = K \delta\left[ E - E_{max}(t)\right]$. In the absence of a fundamental description of the way $E_{max}$ changes in time (it depends on the details of magnetic field amplification and damping) we assume, for $t>T_{ST}$, $E_{max}(t) = E_{M} (t/T_{ST})^{-\alpha}$ ($\alpha>0$), with $E_{M}$ the maximum energy reached at the time $T_{S}$. The value of $\alpha$ is chosen in such a way as to achieve that the maximum energy at the time when the SNR dies out, $\tau_{SNR}$, is $E_{max}(\tau_{SNR})=E_{0}$ where we take $E_{0}=1$ GeV (for protons). The normalization constant $K$ is calculated by integrating $Q(E)$ over $E$:
\be
\int dE Q(E,t) E = \eta(t) \left(\frac{1}{2}\rho u_{sh}^{3} \right) \left( 4 \pi R_{sh}^{2} \right) \to K(t) = \eta(t) \frac{\epsilon_{kin}}{T_{ST}} 
\left( \frac{6}{5}\right)^{3} \frac{1}{E_{M}} \left( \frac{t}{T_{ST}} \right)^{\alpha-1}.
\ee

For reasons that will be clear in the following, we also assume that $\eta(t)=\eta_{0}(t/T_{ST})^{\beta}$, with $\beta\geq 0$.
It follows that the density of CRs from a SNR at location $\vec r_{s}$ is 

$$
n(E,\vec r,t) = \int_{T_{ST}}^{Min[T_{ST}+\tau_{SNR},t-t_s]} dt^{*} \eta(t^{*}) \frac{\epsilon_{kin}}{T_{ST}} 
\left( \frac{6}{5}\right)^{3} \frac{1}{E_{M}} \left( \frac{t^{*}}{T_{ST}} \right)^{\alpha-1} \times 
$$
\be
\delta\left[ E - E_{M}\left( \frac{t^{*}}{T_{ST}} \right)^{-\alpha} \right] {\cal G}_{k}(\vec r, t; \vec r_{s},t^{*}) .
\label{eq:sol}
\ee

The variable $\tau$ in Eq. \ref{eq:green} is clearly $\tau=t-t^{*}-t_{s}$, and the $\delta$-function in Eq. \ref{eq:sol} selects the time $t^{*}=T_{ST}(E_{M}/E)^{1/\alpha}$. One can then write:

$$
n(E,\vec r,t) = \eta_{0} ~\epsilon_{kin} \left( \frac{6}{5}\right)^{3} \frac{1}{\alpha E_{M}^{2}}
\left( \frac{E_{M}}{E} \right)^{2} \left( \frac{E_{M}}{E} \right)^{\frac{\beta}{\alpha}} \times
$$
$$
\frac{1}{\left[ 4 \pi D(E) \tau^{*} \right]^{3/2}} \exp\left[ -\Gamma_{k}^{sp}(E) \tau^{*} \right]
\exp\left[ -\frac{(x-x_{s})^{2}+(y-y_{s})^{2}}{4 D_{k} \tau^{*}}\right] \times
$$
\be
\sum_{n=-\infty}^{+\infty} (-1)^{n} \exp \left[ -\frac{(z-z'_{n})^{2}}{4 D_{k} \tau^{*}}\right],
\label{eq:sol1}
\ee
where 
$$
\tau^{*}=t-t_{s}-T_{ST}\left( \frac{E_{M}}{E} \right)^{\frac{1}{\alpha}}.
$$

From Eq.~\ref{eq:sol} it is clear that the solution is non-zero only for 
$$
1\leq \left( \frac{E_{M}}{E} \right)^{\frac{1}{\alpha}} \leq \rm Min\left[ 1+\frac{\tau_{SNR}}{T_{ST}},\frac{t-t_{s}}{T_{ST}}\right].
$$
The first inequality is satisfied by definition of $E_{M}$. The second inequality reads differently for old and young SNRs. For old SNRs, namely the ones that have completed their evolution before the observation time $t$, the condition reads $E\geq E_{M}\left( 1+\frac{\tau_{SNR}}{T_{ST}} \right)^{-\alpha}$, which is again satisfied by definition for the energies we are interested in, since we required that $E_{M}\left( \frac{\tau_{SNR}}{T_{ST}} \right)^{-\alpha} =E_{0}$. More interesting is the case in which $(t-t_{s})/T_{ST}\leq 1+\tau_{SNR}/T_{ST}$, namely the case of recent SNRs. In this case the spectrum in Eq.~\ref{eq:sol1} is non-zero only for 
$$
E>E_{M} \left( \frac{t-t_{s}}{T_{ST}}\right)^{-\alpha}.
$$
For instance, for $\tau_{SNR}=10^{5}$ years, $E_{M}=3\times 10^{6}$ GeV and $T_{ST}\approx 1000$ years one has $\alpha\approx 3.2$ in order to satisfy $E_M(\tau_{SNR})=E_0=$1GeV. Therefore from a supernova that went off 20,000 years ago, such as Vela, we can only receive particles with energy $E>200$ GeV. It is worth noticing that this fact has nothing to do with the propagation time, which adds to limiting the minimum energy of the particles that can be detected at Earth. The limit discussed here is due instead to the fact that low energy particles could not leave the source even at relatively long times after the beginning of the ST phase. This effect is unlikely to be important for the total spectrum of CRs observed at Earth, which is dominated by the superposition of numerous distant SNRs, but it can potentially be important for the anisotropy signal as we discuss in Paper II. 

One last point that is worth noticing about Eq.~\ref{eq:sol1} is related to the injection spectrum that can be obtained in the approach illustrated above: one can see that if the acceleration efficiency is constant in time ($\beta=0$) the spectrum is always $\propto E^{-2}$ and is totally independent on the spectrum of particles accelerated at the shock. The power law shape of the spectrum is uniquely related to the evolution in time of the SNR and not to the acceleration process at work at the shock. This issue, discussed at length in \cite{Caprioli:2010p133}, is hard to evade and in fact all physical effects that can be added to this approximate picture almost invariably lead to spectra even flatter than $E^{-2}$. Therefore a crucial assumption that is required in order to have injected spectra steeper than $E^{-2}$ is that $\beta>0$. Physically, this implies that the acceleration efficiency is required to increase with time while the SNR gets older. At present it has not been possible to find any physical justification for this requirement.

\section{Simple benchmark results on spectra}\label{sec:simple1}

In this section we provide a benchmark calculation of the spectrum of nuclei expected from discrete sources (SNRs) in the disc of the Galaxy. The full calculation, taking into account the spatial and temporal distribution of SNRs in the Galaxy, will be illustrated in \S~\ref{sec:results}. Here we simply present an analytical calculation that shows the main relevant scaling relations. For this purpose we consider the disc as infinitely thin in the $z$ direction (the sources are all located in a plane with $z_s=0$), and having a radius $R_{d}$. The flux of CRs is calculated at the center of the disc. In a homogeneous model as this is, this does not induce a large error. The spectrum of protons (no spallation) for bursting sources exploding with a rate $\cR$ is easily written as
\be
n_{CR}(E) = \int_{0}^{\infty} d\tau \int_{0}^{R_{d}} dr \frac{2\pi r}{\pi R_{d}^{2}} \frac{N(E)\cR}{\left[4\pi D(E) \tau \right]^{3/2}}\exp \left[ -\frac{r^{2}}{4 D(E) \tau} \right]
\sum_{n=-\infty}^{+\infty} (-1)^{n} \exp \left[ -\frac{(2 n H)^{2}}{4 D(E) \tau} \right].
\label{eq:ncr1}
\ee
Carrying out the integration on $\tau$ first and then on $r$, one easily obtains
\be
n_{CR}(E) = \frac{N(E) \cR}{2 \pi D(E) R_{d}} \sum_{n=-\infty}^{+\infty} (-1)^{n} \left[ \sqrt{1+\left( \frac{2nH}{R_{d}}\right)^{2}} -
\sqrt{\left( \frac{2nH}{R_{d}}\right)^{2}} \right].
\ee
One can check that the sum over $n$ equals $\sim H/R_{d}$ for $H\ll R_{d}$ so that 
\be
n_{CR}(E) = \frac{N(E) \cR}{2 \pi R_{d}^{2}} \frac{H}{D(E)} \equiv  \frac{N(E) \cR}{2 H \pi R_{d}^{2}} \frac{H^{2}}{D(E)} .
\label{eq:ncr}
\ee
The first expression clarifies that the flux of CRs scales with $H/D(E)$, a result that remains valid even in more complex versions of the calculation and in fact found also in propagation codes such as GALPROP \cite{GALPROP2011} and DRAGON \cite{dragon2010}. The second expression in Eq.~\ref{eq:ncr} clarifies that the observed density of CRs is simply equal to the total injection rate $N(E)\cR$ divided by the volume of the Galaxy $2H \pi R_{d}^{2}$, and multiplied by the escape time $H^{2}/D(E)$. 

It is immediately apparent that the diffusion coefficient plays a crucial role in this type of calculations. Since in this paper we do not calculate the $B/C$ ratio or the abundance of radioactive isotopes, we choose to borrow the normalization of $D(E)$ from the results of propagation codes such as GALPROP and DRAGON. They agree on the conclusion that, if the diffusion coefficient is chosen to be in the form 
\be
D(E) = 10^{28} D_{28} \left( \frac{R}{3 GV}\right)^{\delta} \rm cm^{2} s^{-1} 
\label{eq:diff}
\ee
for rigidity $R>3$ GV, the secondary to primary ratios and the abundances of unstable isotopes can be best fit by choosing $D_{28}/H_{kpc}= 1.33$ for $\delta=1/3$ and  $D_{28}/H_{kpc}= 0.55$ for $\delta=0.6$, where $H_{kpc}$ is the height of the halo in units of kpc.

Let us now discuss the case of nuclei, which is somewhat more interesting due to the effect of spallation. The density of CR nuclei of type $k$ in our simple model can be written as
$$
n_{k}(E) = \int_{0}^{\infty} d\tau \int_{0}^{R_{d}} dr \frac{2\pi r}{\pi R_{d}^{2}} \frac{N_{k}(E)\cR}{\left[4\pi D_{k}(E) \tau \right]^{3/2}}\exp\left[-\frac{\tau}{\tau_{sp,k}}\right]\times
$$
\be
\exp \left[ -\frac{r^{2}}{4 D_{k}(E) \tau} \right]\sum_{n=-\infty}^{+\infty} (-1)^{n} \exp \left[ -\frac{(2 n H)^{2}}{4 D_{k}(E) \tau} \right],
\ee
where we defined $\tau_{sp,k}=1/\Gamma_{sp,k}$. The integrals over $\tau$ and $r$ are both carried out analytically in this order, leading to:
$$
n_{k}(E) = \frac{N_{k}(E) \cR}{2 \pi D_{k}(E) R_{d}^{2}} \sqrt{D_{k}(E)\tau_{sp,k}}   \times
$$
\be
\sum_{n=-\infty}^{+\infty} (-1)^{n}\left\{
\exp\left[ -\left(  4 n^{2} \frac{\tau_{esc,k}}{\tau_{sp,k}} \right)^{1/2}\right]-
\exp\left[ - \left(\frac{\tau_{esc,k}}{\tau_{sp,k}}\right)^{1/2} \left(  4 n^{2} + \frac{R_{d}^{2}}{H^{2}}\right)^{1/2}\right]
\right\},
\label{eq:nk}
\ee
where $\tau_{esc,k}(E) = H^{2}/D_{k}(E)$ is the escape time of nuclei of type $k$. The sum in Eq.~\ref{eq:nk} tends to $\sqrt{\tau_{esc,k}/\tau_{sp,k}}$ for $\tau_{esc,k}/\tau_{sp,k}\ll 1$ and to 1 for $\tau_{esc,k}/\tau_{sp,k}\gg 1$. It follows that when spallation is negligible
\be
n_{k}(E) = \frac{N_{k}(E) \cR}{2 \pi R_{d}^{2}} \frac{H}{D_{k}(E)}, ~~~~~  \frac{\tau_{esc,k}}{\tau_{sp,k}} \ll 1
\label{eq:high}
\ee
which is the same solution as for protons (Eq.~\ref{eq:ncr}), showing an energy dependence $n_{k}(E) \propto E^{-\gamma-\delta}$. In the opposite limit (spallation dominant over escape) one has to be careful in defining the effective density experienced by the propagating CRs, which now becomes energy dependent and given by 
\be
n_{gas}(E)=n_{disc}\frac{h}{\sqrt{D(E)\tau_{sp}}}=\frac{n_{disc}^2 h^2 c \sigma_{sp}}{D(E)}\ .
\label{eq:spdens}
\ee
For the CR spectrum one obtains:
\be
n_{k}(E) =  \frac{N_{k}(E) \cR}{2 H \pi R_{d}^{2}} \sqrt{\tau_{sp,k}\tau_{esc,k}}~,  ~~~~~ \frac{\tau_{esc,k}}{\tau_{sp,k}} \gg 1,
\label{eq:low}
\ee
which leads to $n_{k}(E)\propto E^{-\gamma}$, due to the fact that with $n_{gas}$ given by Eq.~\ref{eq:spdens}, $\tau_{sp,k}\propto D(E)$, so that the dependence of the spectrum on the diffusion coefficient cancels out. This implies that at the transition between the two regimes one expects a progressive steepening of the spectrum of nuclei, asymptotically tending to $\delta$ for a spallation dominated regime. This is particularly relevant for heavy nuclei, for which the timescales for spallation and escape become comparable in the TeV range. However, the transition region between the two regimes is rather broad and in general a low energy hardening of the nuclear spectra (with respect to the naive $E^{-\gamma-\delta}$) may be expected even if spallation is not dominant over escape. In the case of He, with our choice of the parameters ($n_{disc}=3cm^{-3}$), $\tau_{sp}=\tau_{esc}$ occurs at $E_{crit}\approx100 GeV/n$, but the spectrum is sensibly flatter than that of H up to around the maximum energy. 

Notice that the disk-like structure of the spatial distribution of sources changes the shape of the spectrum of nuclei compared with the simple leaky-box model predictions used in Ref.~\cite{Horandel:2007p877}.

\subsection{Contribution of a nearby source}\label{sec:nearby}

Here we discuss the relative contribution of an individual source with respect to the diffuse CR spectrum in the Galaxy. For simplicity we limit ourselves to the case of protons, while the generalization to nuclei of arbitrary charge $Z$ is straightforward. The spectrum of the source is assumd to be $\lambda_{N} N(E)$ where $N(E)$ is the average spectrum injected by SNRs, and appearing in Eq.~\ref{eq:ncr}. The CR density contributed by the (bursting) source is 
\be
n_{CR}^{s}(E) \approx \frac{\lambda_{N} N(E)}{\left( 4\pi \lambda_{D} D(E) \tau \right)^{3/2}},
\ee
where the coefficients $\lambda_{N}$ and $\lambda_{D}$ have been introduced to allow for the possibility that the luminosity of the source be somewhat different from the mean luminosity of the CR sources, and that the local diffusion coefficient may be different from that averaged over the entire volume of the Galaxy.

This density needs to be compared with the diffuse density as given by Eq.~\ref{eq:ncr}. The condition that the flux from the source dominates over the diffuse CR flux leads to the requirement:
\be
\frac{\lambda_{N}}{\left( 4\pi \lambda_{D} D(E) \tau \right)^{3/2}} \geq \frac{\cR}{2\pi R_{d}^{2}}\frac{H}{D(E)}.
\ee
With this prescription, the condition that a local source dominates the CR flux reads
\be
\tau \leq 3\times 10^{4} H_{kpc}^{-1} \lambda_{N}^{2/3} \lambda_{D}^{-1} \left( \frac{R}{3 GV}\right)^{-\delta/3} ~ \rm years\ ,
\ee
where we assumed $\cR$=1/(30 yr) and $R_d=15$ kpc.
This result holds within a distance
\be
r \leq \sqrt{4 \lambda_{D} D(E) \tau} \approx 70 \lambda_{N}^{1/3} \left( \frac{R}{3 GV}\right)^{\delta/3} \rm pc
\ee
from the source. Within such distance the density of CR stays roughly constant. 

The two $\lambda$ parameters have been introduced here with a well defined purpose: the luminosity in the form of CRs of a nearby source may be somewhat different from the average one, so that $\lambda_{N}$ allows to consider the possibility that the local source is less powerful ($\lambda_{N}<1$) or more powerful ($\lambda_{N}>1$) than average. For simplicity we kept the spectrum of accelerated particles the same in shape while only the normalization is allowed to change. 

The meaning of $\lambda_{D}$ is more interesting: the diffusion of CRs in the source vicinity can be modified by the propagating particles as a result of the self-generation of turbulence. This effect leads to lowering the diffusion coefficient with respect to average, thereby leading to $\lambda_{D}<1$ (again here we assume for simplicity that only the absolute normalization of $D(E)$ changes while retaining the energy dependence). This effect is even more plausible since, as discussed above, the number density of CRs within a distance $r$ from the source is larger than on average in the Galaxy for a time $\tau$, so that self-generation of waves during this time may reduce the diffusivity of the particles trying to leave the source. For instance if $\lambda_{D}\sim 0.1$ the source contribution dominates on the diffuse CR flux for $\sim 100,000$ years (for $H=3$ kpc). Since the energy dependence of $\tau$ is very weak, this excess is present for a wide range of particle rigidities. The distance $r$ is obviously independent of the value of $\lambda_{D}$. 

\section{Estimate of the fluctuations in the spectrum}\label{sec:simple2}

One of the goals of the present paper is to discuss the implications of diffusive CR propagation from discrete sources on the spectrum measured at Earth. In this section we discuss some basic concepts which help us clarifying the role of fluctuations in the source distribution on the spectrum. In order to quantify these effects we follow here the formalism of Ref. \cite{Lee:1979p1621}.

Based on this formalism, also adopted by \cite{Ptuskin:2006p620}, given  $\cN$ independent sources, each one producing $N_{0}$ particles, the average spectrum these produce at a location $\vec r$ at time $t$ is:
\be
\langle n_{CR} (\vec r,t) \rangle =  N_{0} \cN \int d^{3}\vec r' dt' P(\vec r',t') \cG (\vec r,t;\vec r',t'),
\ee
where $P(\vec r', t')$ is the probability of having a source with space-time coordinates $(\vec r', t')$ and $\cG(\vec r, t; \vec r', t')$ is the Green function for transport of particles from $(\vec r', t')$ to $(\vec r, t)$.
For the fluctuations in the measured spectrum one finds \cite{Lee:1979p1621}:
$$
\langle \delta n_{CR}(\vec r,t) \delta n_{CR} (\vec r',t') \rangle = 
$$
\be
= N_{0}^{2} \cN \int d^{3}\vec r'' dt'' P(\vec r'',t'') \cG (\vec r,t;\vec r'',t'')  \cG (\vec r',t';\vec r'',t'') -
\frac{1}{\cN} \langle n_{CR}(\vec r,t)\rangle \langle n_{CR}(\vec r',t')\rangle.
\ee 

In our case, the probability distribution, as already used in Eq.~\ref{eq:ncr1}, is 
\be
P(\vec r,t) = \frac{\cR}{\pi R_{d}^{2}}\ ,
\ee
where, by definition, $\cR=\cN/\Delta t$ with $\Delta t\to\infty$ the time over which the averages are calculated. For simplicity, we carry out the calculation for the simple case in which the sources are uniformly distributed in an infinitely thin disc with radius $R_{d}$, and assume for our estimates that the spectrum and its fluctuations are measured at the center of the disc, $r=0$. Moreover we limit our simple estimate to protons, hence ignoring spallation terms in the Green function. With these assumptions, the integral over the spatial coordinates reduces to an integral over cylindrical radius. The double integral over space and time diverges unless a lower bound is imposed on one of the two coordinates, either a minimum distance from the closest source, $R_{min}$, or a minimum time for receiving its contribution, $T_{min}$: the two are obviously correlated, with $R_{min}\sim\sqrt{4 D T_{min}}$ and we choose to impose a $T_{min}$ (see \cite{mertsch2011} for a more extended discussion of this effect). The correlation function evaluated in the same spatial location is then:
\be
\langle \delta n_{CR}(\vec r,t) \delta n_{CR} (\vec r,t) \rangle \approx \frac{\cR N_{0}^{2}}{32 \pi^{3} R_{d}^{2} D(E)^{2} T_{min}},
\ee
where we assumed that $H^{2}/(D(E)T_{min})\gg 1$. The strength of the fluctuations is therefore:
\be
\delta_{spec} (E) = \frac{\langle \delta n_{CR}\delta n_{CR}\rangle^{1/2}}{n_{CR}(E)} = \frac{1}{2^{3/2} \pi \left( \frac{\cR}{\pi R_{d}^{2}}\right)^{1/2} H T_{min}^{1/2}}.
\ee

This result shows the strong dependence of $\delta_{spec}$ (in terms of normalization and in terms of energy dependence) on the cutoff time $T_{min}$, which we imposed in order to avoid the divergence in the integral over $\tau$. One can see that if $T_{min}$ is naively chosen as a given number, then $\delta_{spec}$ turns out to be independent of energy. On the other hand, from the physical point of view, the time $T_{min}$ should carry information about the closest, most recent sources around the observer. For instance the time $T_{min}$ could be interpreted as the time over which one source goes off within a distance from Earth such that CRs from that source reach us within a time $T_{min}$. This condition is expressed as
\be
\frac{\cR T_{min}}{\pi R_{d}^{2}} \left[4 \pi D(E)T_{min}\right] = 1 \to T_{min} = \left[\frac{4 \cR D(E)}{R_{d}^{2}}\right]^{-1/2}.
\ee
With this prescription one obtains \cite{Ptuskin:2006p620}:
\be
\delta_{spec} (E) = \frac{D(E)^{1/4}}{2 \pi^{3/4} H \left( \frac{\cR}{\pi R_{d}^{2}}\right)^{1/4}}\propto E^{\frac{\delta}{4}} .
\ee
Since the diffusion coefficient is chosen as in Eq.~\ref{eq:diff}, the expression above can be rewritten as:
\be
\delta_{spec} (E) \approx 0.03 H_{kpc}^{-3/4} \cR_{30}^{-1/4}R_{d,15kpc}^{1/2} \left( \frac{E}{3 GeV} \right)^{\delta/4},
\label{eq:deltasp}
\ee
where we have used $D_{28}/H_{kpc}\sim 1$ (the exact value being irrelevant since the dependence of $\delta_{spec}$ on this ratio is very weak).

The energy dependence of the spectral fluctuations is very weak for the standard values of $\delta$. At energies $E\sim 10^{5}$ GeV one can expect fluctuations at the level of $\sim 5\%$ for $\delta=1/3$ and at the level of $10\%$ for $\delta=0.6$. In practice, as we show below, the fluctuations may be somewhat larger because of the combined effect of all nuclei and the occasional dominance of some local source contributing most of the CR flux at Earth for some time (see discussion in \S \ref{sec:nearby}). This latter scenario is most likely to happen for $\delta\sim 0.6$. 

\section{Results for realistic distributions of SNRs}\label{sec:results}

In this section we describe our more realistic calculations of spectra and chemical composition, taking into account the discreteness in space and time of sources in the Galaxy. First of all let us describe our model of CR diffusion in the Galaxy. We assume that the diffusion region has a height $H$ above and below the disc, with the latter assumed to have half-width $h$. In the radial direction the diffusion region is not bounded. As we discuss below, this assumption is not a serious limitation thanks to the form of the SNR spatial distribution. The diffusion coefficient in the whole region is assumed to be spatially constant (the same assumption as in propagation codes such as GALPROP) and to depend on rigidity alone, as in Eq.~\ref{eq:diff}. As suggested by the results of GALPROP, as well as of other propagation codes, a good fit to the whole set of available CR data is obtained by taking $D_{28}/H_{kpc}\simeq 1$ (more precisely $D_{28}/H_{kpc}=1.33$ for $\delta=1/3$ and $D_{28}/H_{kpc}=0.55$ for $\delta=0.6$), therefore we adopt this relative normalization of diffusion coefficient and size of the halo. 

We adopt two different models of source distributions. The first one (cylindrical model hereafter) simply accounts for the average radial and azimuthal distribution of sources in the Galaxy, as discussed for instance in \cite{Case:1996p1635}. The other model is instead an attempt at taking into account the spiral distribution of sources (hereafter spiral model), for which we adopt the formalism of \cite{FaucherGiguere:2006p1609}.

The sources are assumed to be distributed in the Galaxy according to the following function \cite{Case:1996p1635}:
\be
f(r)=\frac{A}{R_{\odot}^{2}}\left( \frac{r}{R_{\odot}}\right)^{2} \exp \left[ -\beta\frac{r-R_{\odot}}{R_{\odot}}\right],
\label{eq:radial}
\ee
where $\beta=3.53$ for $R_{\odot}=8.5$ kpc. 

The constant $A$ is determined from the normalization condition:
\be
\int_{0}^{\infty} dr~2\pi r f(r) = 1 \to A=\frac{\beta^{4}\exp(-\beta)}{12 \pi}.
\ee
The SNR distribution in the $z$ direction is assumed to be 
\be
f(z) = \frac{A_{z}}{z_g}\exp\left[ -\frac{\abs{z}}{z_g}\right],
\ee
where $A_{z}=1$ is again derived from the normalization condition. 

In the cylindrical model, the positions of the sources are chosen by drawing at random values of $r$ and $z$ from the distributions above. The $x$ and $y$ coordinates are chosen by generating a random angle $0\leq \phi \leq 2\pi$ and using the given value of $r$. Supernovae are generated at a rate $\cR=(30-100~ \rm yr)^{-1}$, and the generation of new sources is continued until a time span much larger than $H^{2}/D(E)$ is covered, in order to make sure that the stationary solution has been reached at the lowest rigidities of interest for us ($\sim 1$ TV).

In the {\it spiral model}, the procedure we adopt is similar to that introduced in \cite{FaucherGiguere:2006p1609}. The generation of the position in the $z$ direction is the same as above, therefore we will not discuss it any further. A radial coordinate $\tilde r$ is drawn at random from the average distribution in Eq.~\ref{eq:radial}; at this point a random natural number is chosen between 1 and 4, with a flat distribution. This number identifies the arm in which the supernova is localized (Norma, Carina-Sagittarius, Perseus, Crux-Scutum, as in Table 1). At this point an angular position along the arm is associated to the SNR, following the relation:
\be
\theta(r) = K \log \left(\frac{r}{r_{0}}\right) + \theta_{0}.
\ee
The parameters $K$, $r_{0}$ and $\theta_{0}$ are reported in Table 1 (notice that the values of $\theta_{0}$ are different from those in Table 2 of \cite{FaucherGiguere:2006p1609}, simply because the axes are rotated by $\pi/2$ with respect to their choice). The Sun is located at $(x,y,z)=(8.5 kpc,0,0)$.  

\begin{table}
  \centering
\caption{\label{table1} Parameters of Galactic arms.}
\vspace{1ex}
\begin{tabular}{c|*{3}{|@{\hspace{1em}}c@{\hspace{1em}}}}
\hline \hline
  \rule{0pt}{4ex}
  arm number/name &  $K(rad)$ & $r_{0}(kpc)$ & $\theta_{0}(rad)$ \\
\hline \hline
\rule[-2ex]{0pt}{6ex}
1: Norma & 4.25 & 3.48 & 0 \\\hline
\rule[-2ex]{0pt}{6ex}
2: Carina - Sagittarius & 4.25 & 3.48 & $\pi$ \\\hline
\rule[-2ex]{0pt}{6ex}
3: Perseus & 4.89 & 4.90 & 2.52 \\\hline
\rule[-2ex]{0pt}{6ex}
4: Crux - Scutum & 4.89 & 4.90 & -0.62 \\\hline
\hline \hline
\end{tabular}
\end{table}

\begin{figure}[t]
\centering\leavevmode
\includegraphics[width=2.9in,angle=0]{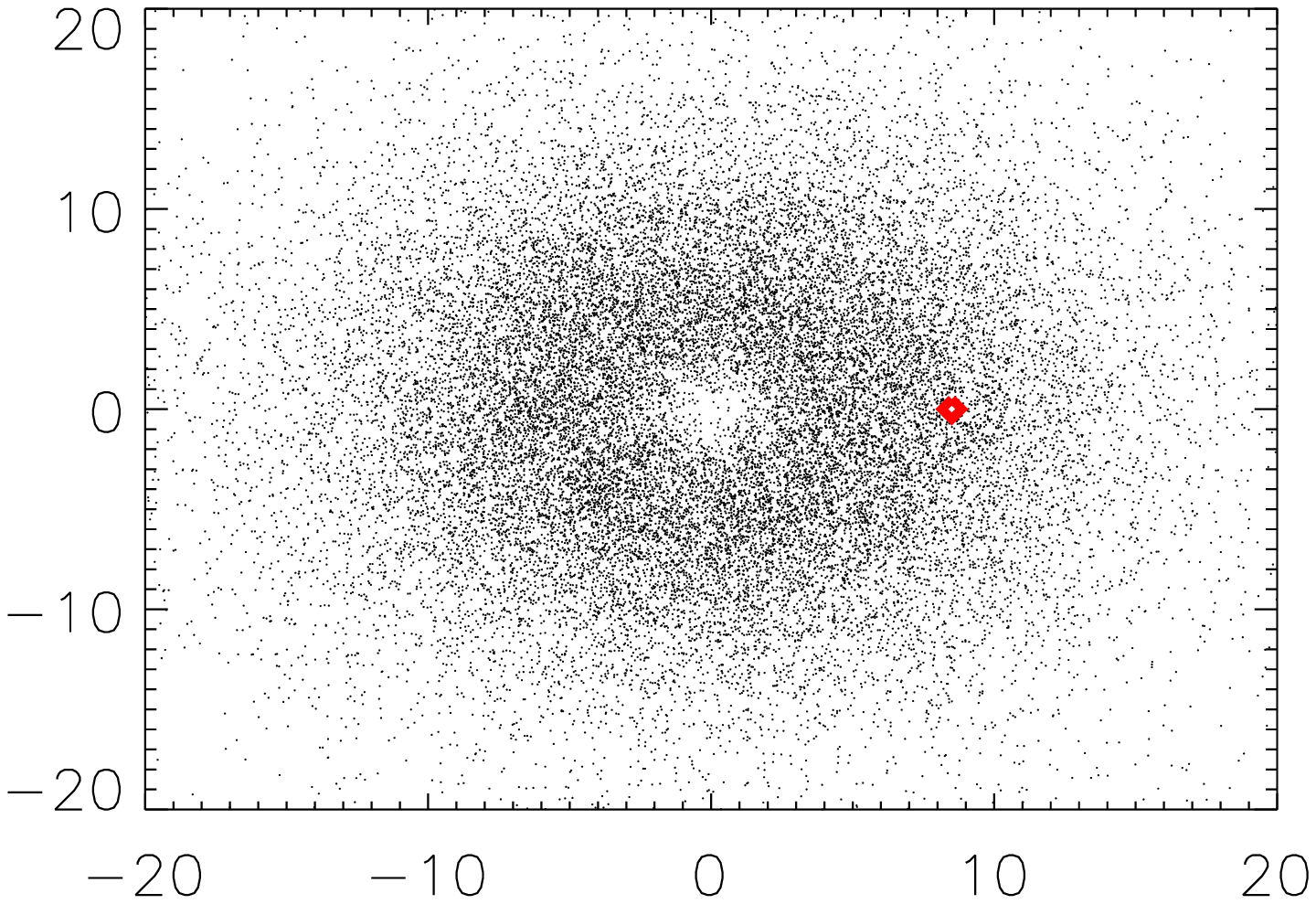}
\includegraphics[width=2.9in,angle=0]{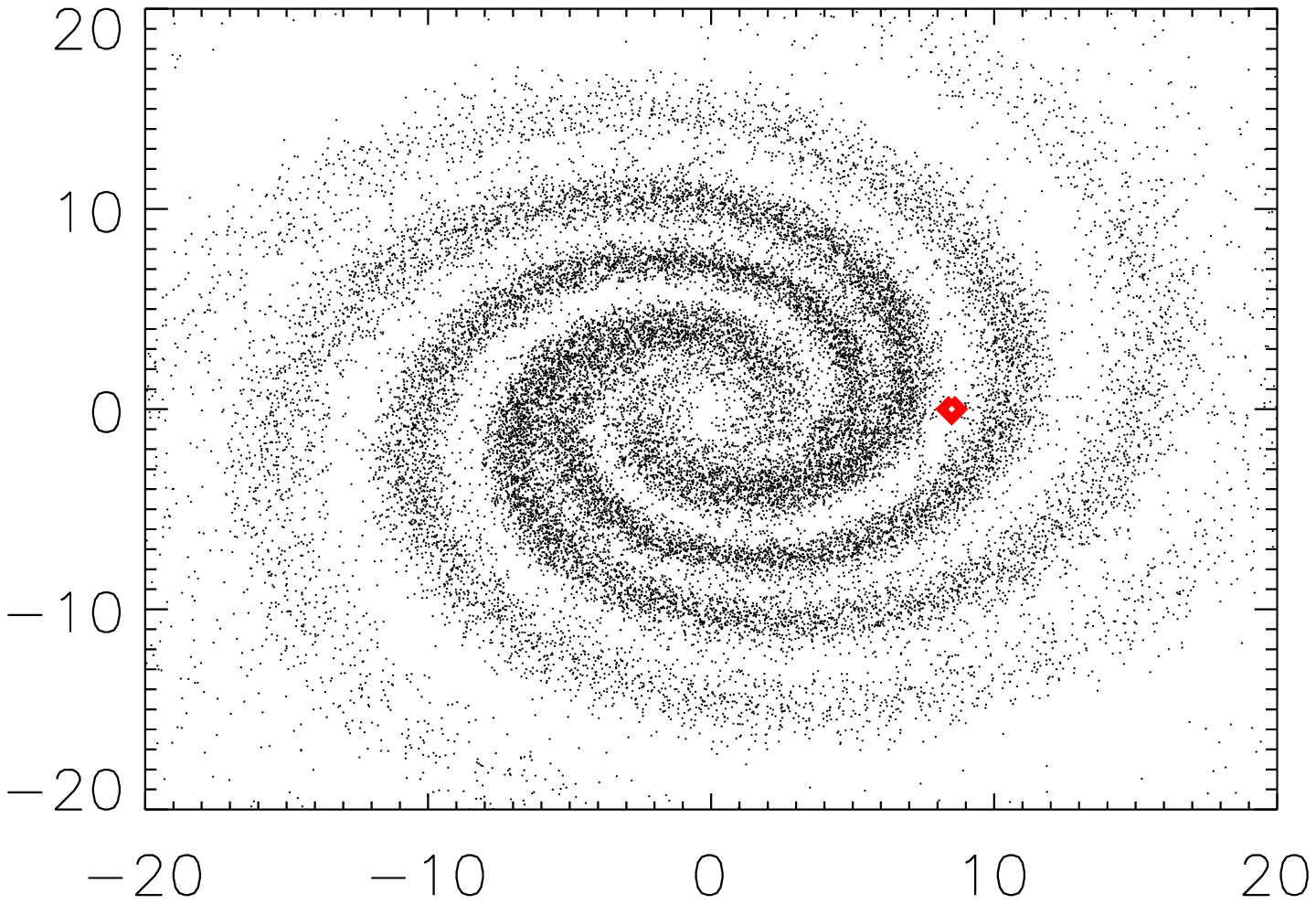}
\caption{A face on view of the spatial distribution of SNRs in the Galaxy in the two models: cylindrical in the left panel and spiral on the right. About $3 \times 10^4$ sources are shown in each panel. Units are in $kpc$ and the position of the Sun is marked by a thick (red) symbol.} 
\label{fig:space}
\end{figure}

Following the prescription of Ref. \cite{FaucherGiguere:2006p1609}, we blur the angle $\theta(r)$ by $\theta_{corr}\exp(-0.35 \tilde r/kpc)$, where $\theta_{corr}$ is chosen from a flat random distribution between 0 and $2\pi$. Similarly the radial coordinate is also blurred by choosing a {\it new} value from a normal random distribution with mean $\tilde r$ and variance $0.07\tilde r$. A pictorial illustration of the two scenarios is shown in Fig. \ref{fig:space} where we show the distribution of $\sim 30,000$ SNRs in the cylindrical model (left) and the spiral model (right). The position of the Sun is illustrated by the thick (red) symbol. 

For each source the spectrum of CRs (protons, He nuclei, CNO nuclei, Mg-Al-Si nuclei and Fe nuclei) at the Earth is calculated using the appropriate Green functions, as described in \S~\ref{sec:green}. For each realization of the source distribution in the Galaxy we also compute the chemical composition of CRs, as derived from the superposition of the flux of different chemicals. The efficiencies of acceleration of nuclei are calculated {\it a posteriori} from requiring a fit to the available spectra in the $TeV$ region. The calculations are carried out for different choices of the propagation parameters. We account for spallation as discussed in \S~\ref{sec:green}, with the average gas density in the propagation volume taken as $n_{gas} = n_{disc}(h/H)$. Notice that with this set of prescriptions the grammage traversed by CRs with rigidity $R$ reads
\be
x(R) = n_{gas} c m_{p} \frac{H^{2}}{D(R)} = n_{disc} h c m_{p} \frac{H}{D(R)} \approx x_{0} \left( \frac{n_{disc}}{1~\rm cm^{-3}}\right)  \left( \frac{R}{3~GV}\right)^{-\delta} \rm g~cm^{-2},
\label{eq:grammage}
\ee 
which only depends on the density of gas in the Galactic disc and not on the spatial extent of the halo, once $H_{kpc}/D_{28}\sim 1$ is used. In our calculations we adopt $n_{disc}=3~cm^{-3}$, which leads to $x_{0}\simeq 9~\rm g~cm^{-2}$ for $\delta=1/3$ and $x_{0}\simeq 4.3 ~\rm g~cm^{-2}$ for $\delta=0.6$. It is worth stressing that although Eq.~\ref{eq:grammage} is normalized at rigidity $3$ GV, all of our calculations refer to much higher energies, in the TeV range.

\subsection{Spectra and chemical composition for the cylindrical model}\label{sec:res_cyl}

In this section we discuss many results related to the dependence of the spectra and chemical composition on the diffusion coefficient, on discreteness of the sources in space and time and the size of the halo for the cylindrical model of source distribution.

\begin{figure}[t]
\centering\leavevmode
\includegraphics[width=2.8in,angle=0]{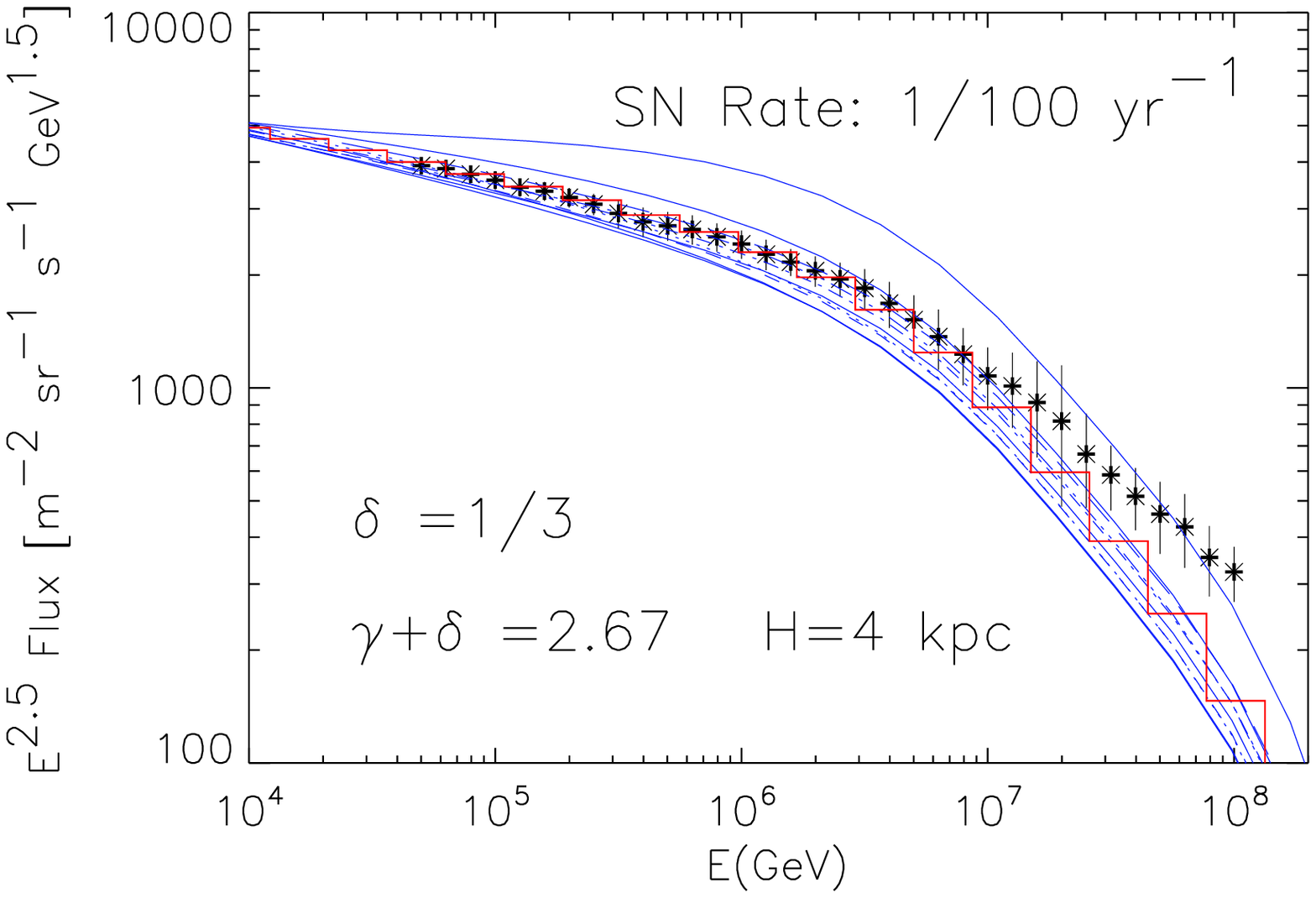}
\includegraphics[width=2.8in,angle=0]{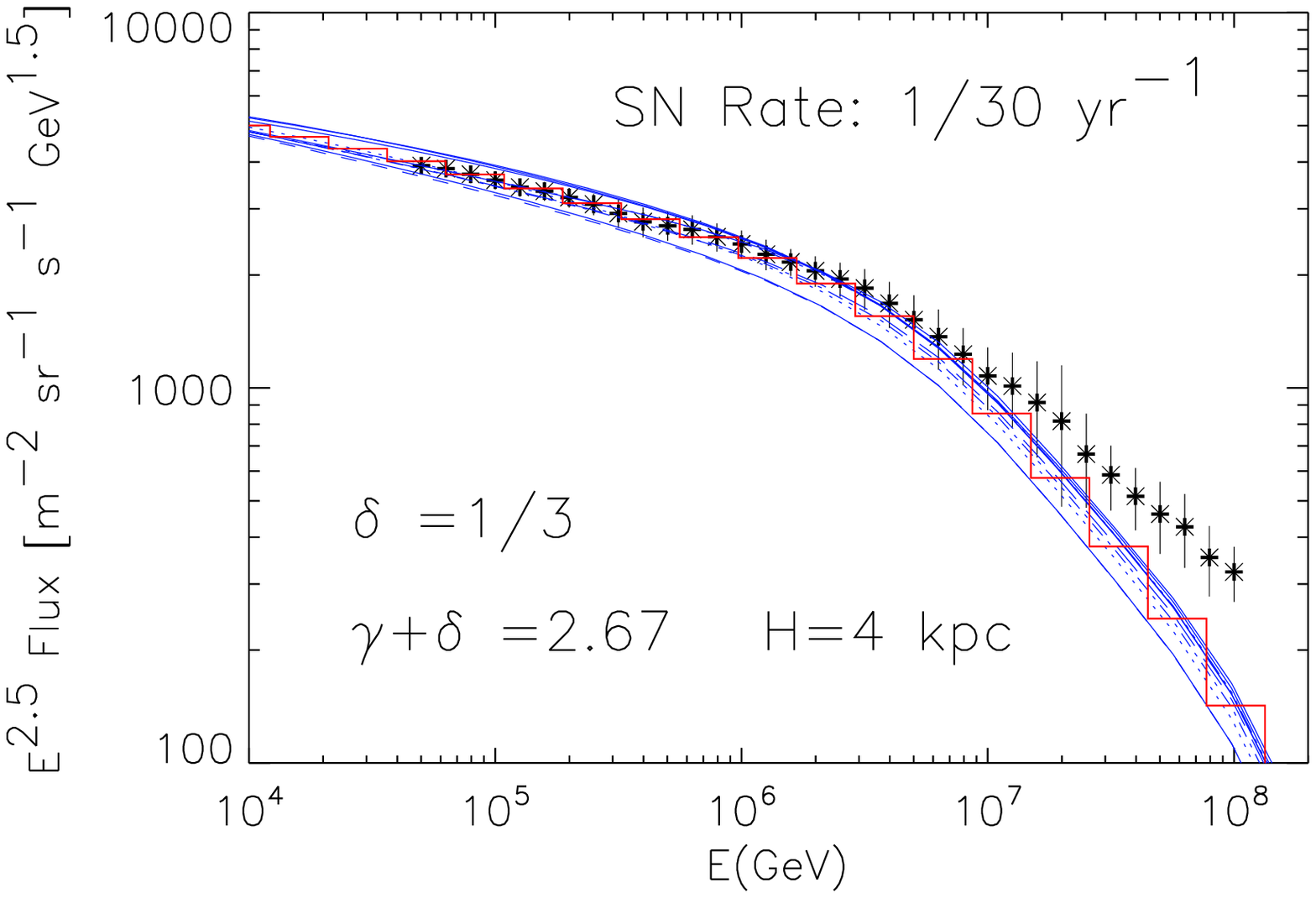}
\caption{All-particle CR spectrum for ten random realizations of sources in the cylindrical model, assuming $\delta=1/3$, a SN rate $R=1/100 yr^{-1}$ ($R=1/30 yr^{-1}$) on the left (right). The halo size is $H=4$ kpc. The injection spectrum of all chemicals is assumed to have a slope (below the cutoff) such that $\gamma+\delta=2.67$.}  
\label{fig:varyR}
\end{figure}

In Fig.~\ref{fig:varyR} we illustrate the all-particle spectrum obtained in 10 realizations of source distributions in the cylindrical model, using $\delta=1/3$, $H=4$ kpc and a rate of supernova explosions in the Galaxy $\cR=1/100 yr^{-1}$ (left) and $\cR=1/30 yr^{-1}$ (right). In both cases we impose that the slope $\gamma$ of the injection spectrum is related to $\delta$ through $\gamma+\delta=2.67$, where 2.67 turns out to be the slope of the data provided in Ref.~\cite{Horandel:2004p1543} as {\it average} among all experiments (dots with error bars in Fig.~\ref{fig:varyR}). The red, staircase line, represents the average spectrum resulting from the 10 random realizations. 

As expected, the random distribution of SNRs in space and time leads to fluctuations in the all-particle spectrum. One aspect that can be immediately noticed is that the strength of the fluctuations is sensibly reduced in the case with a higher SN rate ($\cR=1/30 yr^{-1}$), as could be expected based on the simple estimate in Eq.~\ref{eq:deltasp}. Though relatively small, the effect of fluctuations is sufficient to induce a visible wiggle in the global shape of the all-particle spectrum.

All spectra present a pronounced knee which compares well with the data and is the consequence of the rigidity dependent nature of the acceleration in the sources. In order to provide a reasonably good fit to the data one has to assume a maximum energy in the proton spectrum $E_{max}=6\times 10^{6}$ GeV, not far from that obtained in non-linear theories of particle acceleration at SNR shocks \cite{Blasi:2007p144}. This number can however reasonably be smaller in more realistic models since the spectrum of escaping particles in general does not have an exponential cutoff, due to the superposition of advected particles and particles escaping from upstream at all times throughout the history of the remnant \cite{Caprioli:2008p123}. The fact that our model spectrum underpredicts the data at the highest energies is most likely related to the effect of extragalactic cosmic rays, as we discuss below.

\begin{figure}[t]
\centering\leavevmode
\includegraphics[width=4.8in,angle=0]{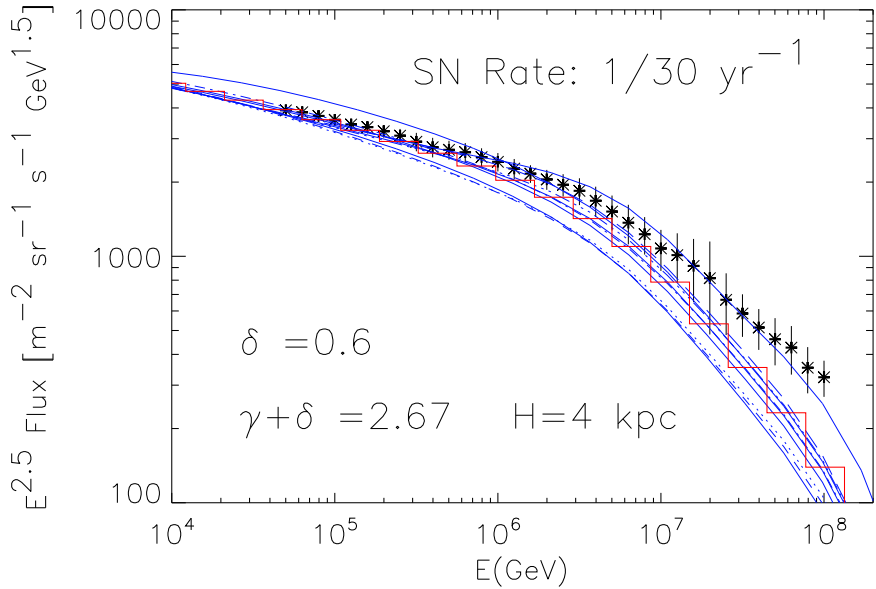}
\caption{All-particle CR spectrum for ten random realizations of sources in the cylindrical model, assuming $\delta=0.6$, a SN rate $R=1/30 yr^{-1}$ and a halo with size $H=4$ kpc. The injection spectrum of all chemicals is assumed to have a slope (below the cutoff) such that $\gamma+\delta=2.67$.}  
\label{fig:H2delta06R30}
\end{figure}

The fluctuations are much more pronounced in the case of faster diffusion ($\delta=0.6$), as illustrated in Fig.~\ref{fig:H2delta06R30}, an effect that could be expected based on Eq.~\ref{eq:deltasp}. Independent of the shape of the average spectrum, one can clearly see that the spectra of individual realizations are in general rather wiggly and can fit the observations only in a rather loose sense. The bumpiness of the spectrum is the result of the importance of nearby and recent supernova events. These realizations also lead to exceedingly large anisotropy as we discuss in Paper 2. Both aspects, however, bumpiness and anisotropy, are likely to hint to the breaking of the diffusion approximation for CRs coming from nearby sources. 

Using Eq.~\ref{eq:diff} for the diffusion coefficient and recalling that $D(E)$ can also be written in terms of the scattering length $\lambda(R)$ as $D(R)=(1/3) c \lambda (R)$, one easily obtains 
\be
\lambda(R) \approx 10^{18} H_{kpc} \left( \frac{R}{3 GV}\right)^{\delta} \rm cm,
\ee
where for simplicity of presentation we assumed $D_{28}/H_{kpc}=1$, independent of $\delta$.

The diffusion approximation for an individual source is valid only if the source is located at a distance $d\gg \lambda$, but at $R=10^{5}$ GV and with $H=4$ kpc, $\lambda\approx 380$ pc for $\delta=0.6$. In other words at high energy one should be very careful in treating the propagation from nearby sources in order to avoid incorrect or artificial conclusions on the spectrum and anisotropy. 

\begin{figure}[t]
\centering\leavevmode
\includegraphics[width=4.8in,angle=0]{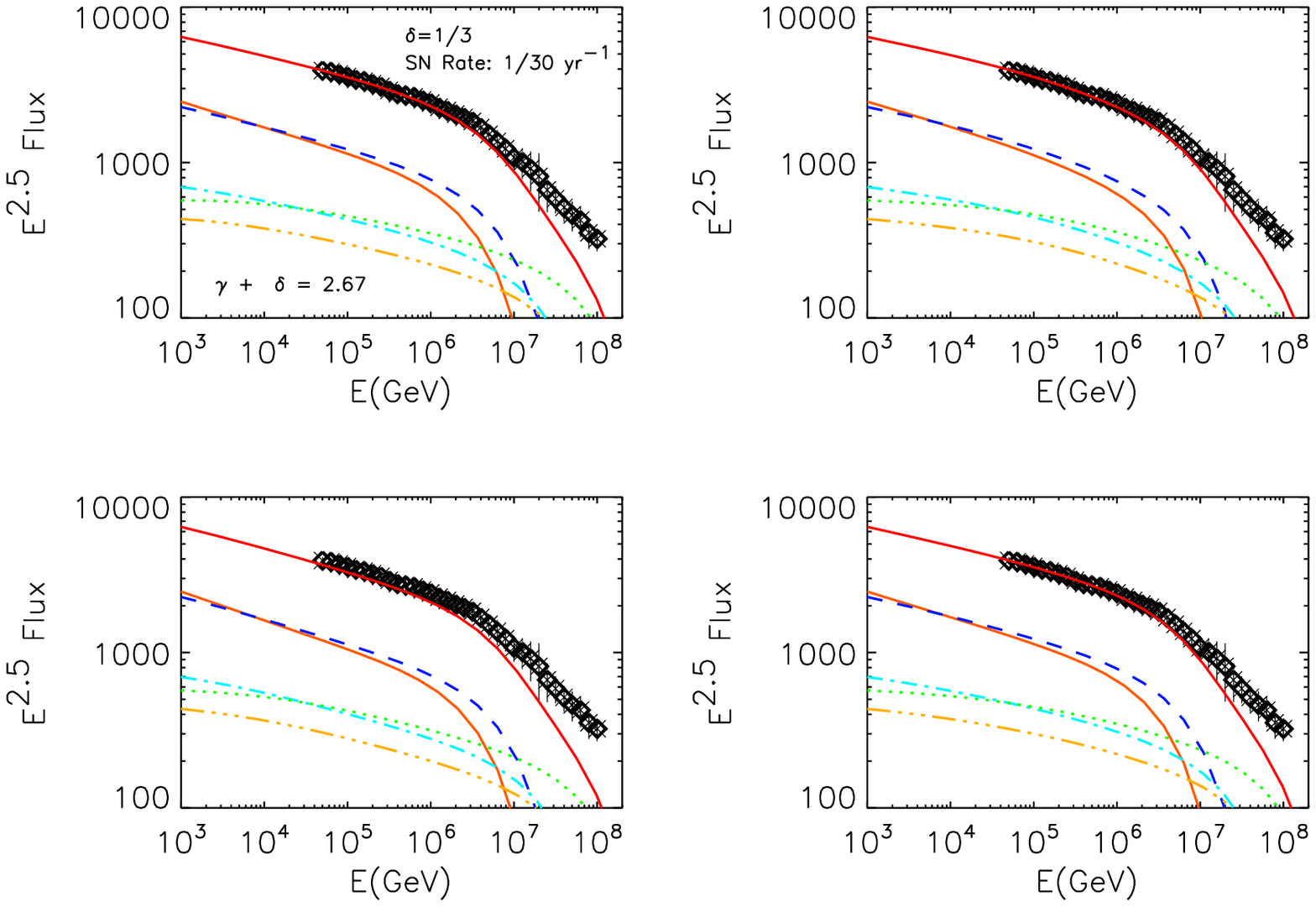}
\caption{Spectra of chemicals in the cylindrical model, for $\delta=1/3$, a SN rate $\cR=1/30 yr^{-1}$ and a halo with size $H=4$ kpc. The injection spectrum of all chemicals is assumed to have slope (below the cutoff) such that $\gamma+\delta=2.67$. The different curves are: solid (orange) for protons, dashed (blue) for He, dot-dashed (cyan) for CNO, dot-dot-dot-dashed (yellow) for Mg-Al-Si, and, finally, dotted (green) for Fe. Each of the four different panels shows the results for a different random realization of the sources' distribution.}  
\label{fig:H2delta033chems}
\end{figure}

It is interesting to discuss the spectra of individual elements for the different realizations. We illustrate our results for $\delta=1/3$ for four realizations in Fig.~\ref{fig:H2delta033chems}. The different lines refer to protons (solid/orange), He (dashed/blue), CNO (dot-dashed/cyan), MgAlSi (dor-dot-dot-dashed/yellow) and Fe (dotted/green). The red solid line is the all-particle spectrum compared with data. 

For $\delta=1/3$ the fluctuations are rather small and a good fit to the all-particle spectrum can always be obtained. The most important feature shown by these curves is the flattening of the spectra of chemicals at lower energies, which results from spallation. In particular the spectrum of He appears to be flatter than the spectrum of protons up to $\sim 10^{5}$ GeV. This prediction is observationally consistent with recent CREAM \cite{Ahn:2010p624,Ahn:2009p1595} and PAMELA data, and implies that at the knee He dominates over protons, even if by a small amount. It is very important to realize that the spectral flattening does not imply that spallation dominates over escape from the Galaxy. The latter condition occurs at lower energies when the escape times become larger. For the energies we are interested in, the flattening is simply induced by the secular action of spallation during the propagation time, and is certainly more pronounced for elements heavier than He, as Fig.~\ref{fig:H2delta033chems} clearly shows. In our calculations the slope of the proton spectrum is $\sim 2.67$ but that of the He spectrum is $\sim 2.6$ (the difference in slope is $\sim 0.05-0.07$ depending on realization). As shown in Eq.~\ref{eq:low}, if spallation were dominant the change of slope would be $\delta\sim 0.33$. These numbers should be compared with the slopes of the proton and He spectra as observed by CREAM, that are $2.66\pm 0.02$ and $2.58\pm 0.02$ \cite{Ahn:2010p624} respectively. 

\begin{figure}[t]
\centering\leavevmode
\includegraphics[width=4.8in,angle=0]{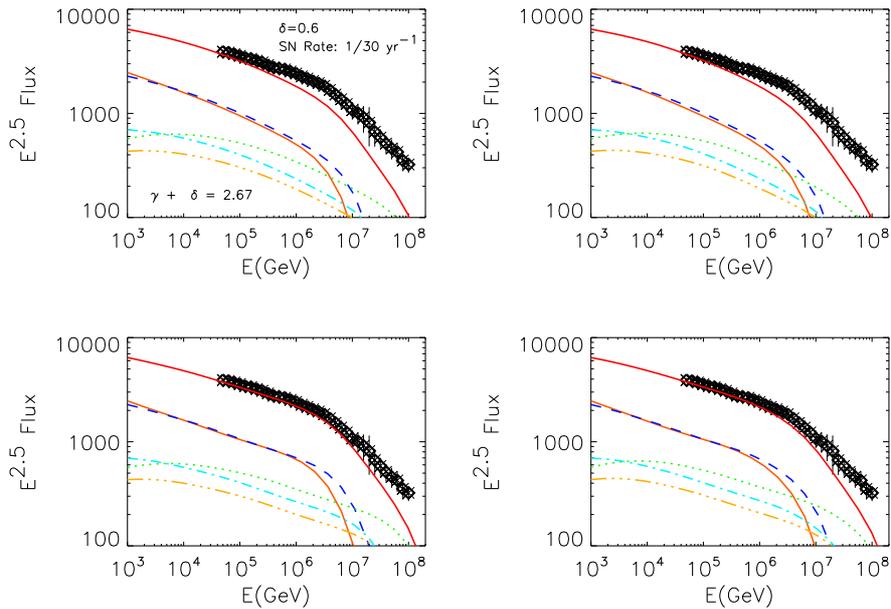}
\caption{Same as for Fig.~\ref{fig:H2delta033chems} but for $\delta=0.6$.}  
\label{fig:H2delta06chems}
\end{figure}

The case $\delta=0.6$ is somewhat different, as shown in Fig.~\ref{fig:H2delta06chems}. As stressed above, the fit to the all-particle spectrum is not that solid and depends on the specific realization. Moreover, the hardening in the He spectrum is not evident and in fact the He and proton spectra are mostly parallel in the realizations shown here. Nearby sources induce wiggles and occasionally give rise to concave spectra. These are not to be confused with the ones predicted by non linear diffusive shock acceleration. 

It is important to keep in mind that the spallation-based explanation of the flatter He spectrum compared with the H spectrum relies on the requirement of a large enough grammage at energies above $200$ GeV/nucleon or so. At these energies the B/C ratio is not very constraining. On the other hand, trying to obtain a comprehensive fit of B/C and particle spectra at lower energies is also problematic, since the spectra of protons and He both show a sharp hardening at $200$ GeV/nucleon, that can, equally well, be the result of local sources or be due to a reduction of the scattering rate of particles (larger diffusion coefficient). The first explanation would not affect the global B/C ratio, while the latter would indeed affect it, since it results in a decreased grammage at low energies. In this sense the origin of the harder He spectrum should be investigated together with the origin of the change of spectral slope seen for all species at $200$ GeV/nucleon.

From the point of view of energetics the two scenarios with $\delta=1/3$ and $\delta=0.6$ are appreciably different. On average the case $\delta=1/3$ leads to require an efficiency of acceleration (for protons) of the order of $\sim 6-7\%$ for $\cR=1/30 yr^{-1}$. For $\delta=0.6$ the efficiency of proton acceleration must be $\sim 10\%$. The fact that (especially for $\delta=1/3$) the acceleration efficiency for individual SNRs is so small might be a possible reason why it is very difficult to detect the gamma ray emission from pion production and decay. It can also provide a useful hint to the development of non linear theories of shock acceleration: on one hand the acceleration efficiency is not required to be high, on the other it has to be large enough to induce the magnetic field amplification necessary for particle acceleration up to the knee region. 

\begin{figure}[t]
\centering\leavevmode
\includegraphics[width=2.8in,angle=0]{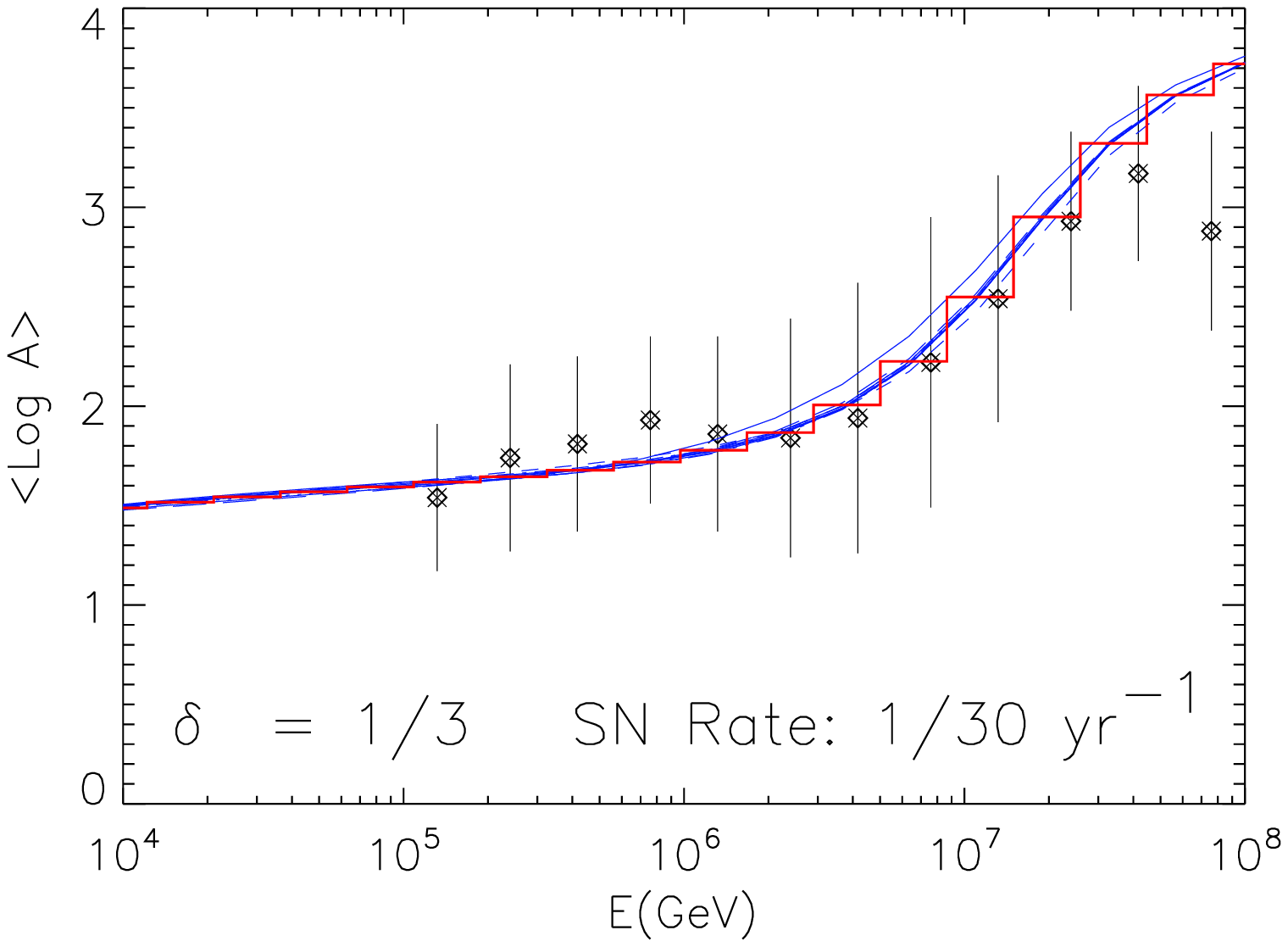}
\includegraphics[width=2.8in,angle=0]{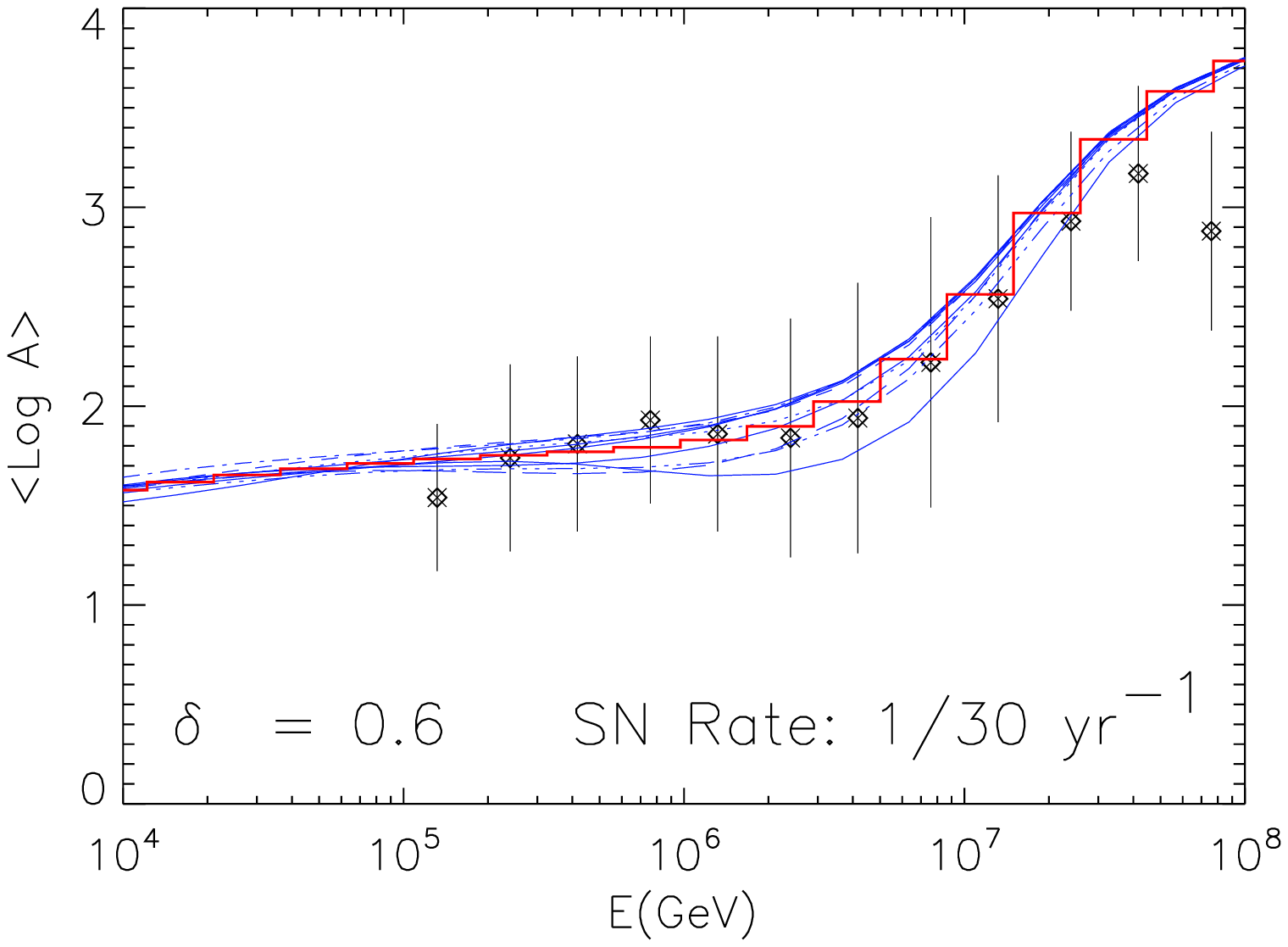}
\caption{Mean Log of the mass of CRs for ten random realizations of the distribution of SNRs in the cylindrical model, assuming $\delta=1/3$ ($\delta=0.6$) on the left (right) and a halo with size $H=4$ kpc. The red staircase line is the average over the random realizations.}  
\label{fig:H2compo}
\end{figure}

A summary of the results on chemical composition is provided by the mean value of the logarithmic mass as a function of energy. We plot this function in Fig.~\ref{fig:H2compo} for $\delta=1/3$ (left) and $\delta=0.6$ (right). The results of our calculations are compared with the data compiled in Ref.~\cite{Horandel:2007p877}. Unfortunately the large error bars in the data do not allow to add information to what we have already pointed out above. The behaviour of the data in the highest energy bins might be interpreted as suggestive of the need for a lighter component, possibly contributed by extragalactic cosmic rays.

\subsection{Spectra and chemical composition for the spiral model}\label{sec:res_spiral}

\begin{figure}[t]
\centering\leavevmode
\includegraphics[width=4.8in,angle=0]{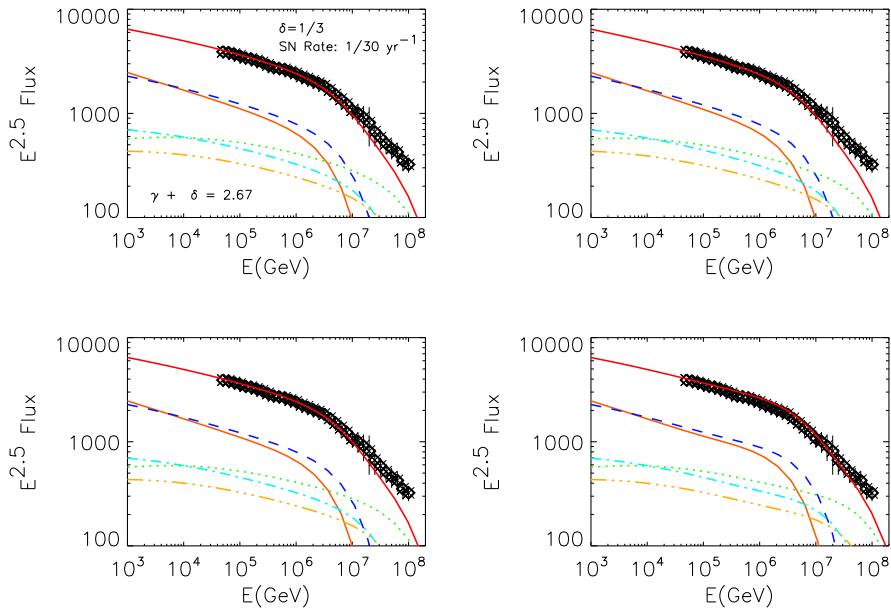}
\caption{Spectra of chemicals in the spiral model, for $\delta=1/3$, a SN rate $\cR=1/30 yr^{-1}$ and a halo with size $H=4$ kpc. The injection spectrum of all chemicals is assumed to have slope (below the cutoff) such that $\gamma+\delta=2.67$.}  
\label{fig:SpH2delta033g267chems}
\end{figure}

In this section we discuss the implications of the spiral distribution of SNRs in the Galaxy for the spectrum and chemical composition of cosmic rays observed at Earth. As shown in Fig.~\ref{fig:SpH2delta033g267chems}, again a very good fit to the measured all-particle spectrum can be obtained adopting the same conditions as for the cylindrical model, namely $\gamma+\delta=2.67$. In this case, however, an interesting dependence of the results on the size of the magnetized halo arises.

From Eq.~\ref{eq:grammage} we see that the grammage depends very weakly on $H$ because of the condition $H_{kpc}/D_{28}\sim 1$. This implies that changing (decreasing or increasing) $H$ has no appreciable consequences in the case of the cylindrical model, where the source density around the Sun is roughly constant (within a distance of order $\sim H$). This conclusion remains approximately true also in the spiral model, although here changing $H$ leads to a change in the volume (and number of arms) contributing a flux at Earth. Now, by taking a look at Fig.~\ref{fig:space}, it is immediate to notice that the position of the Sun is right in between two Galactic arms, so that most of the sources are some distance away. Nuclei that have to be transported from the arm to the Sun suffer spallation and their depletion at low energy is not compensated by the contribution of nearby sources. This is exactly the situation in which changing the volume of contributing sources might make a difference.

\begin{figure}[t]
\centering\leavevmode
\includegraphics[width=2.8in,angle=0]{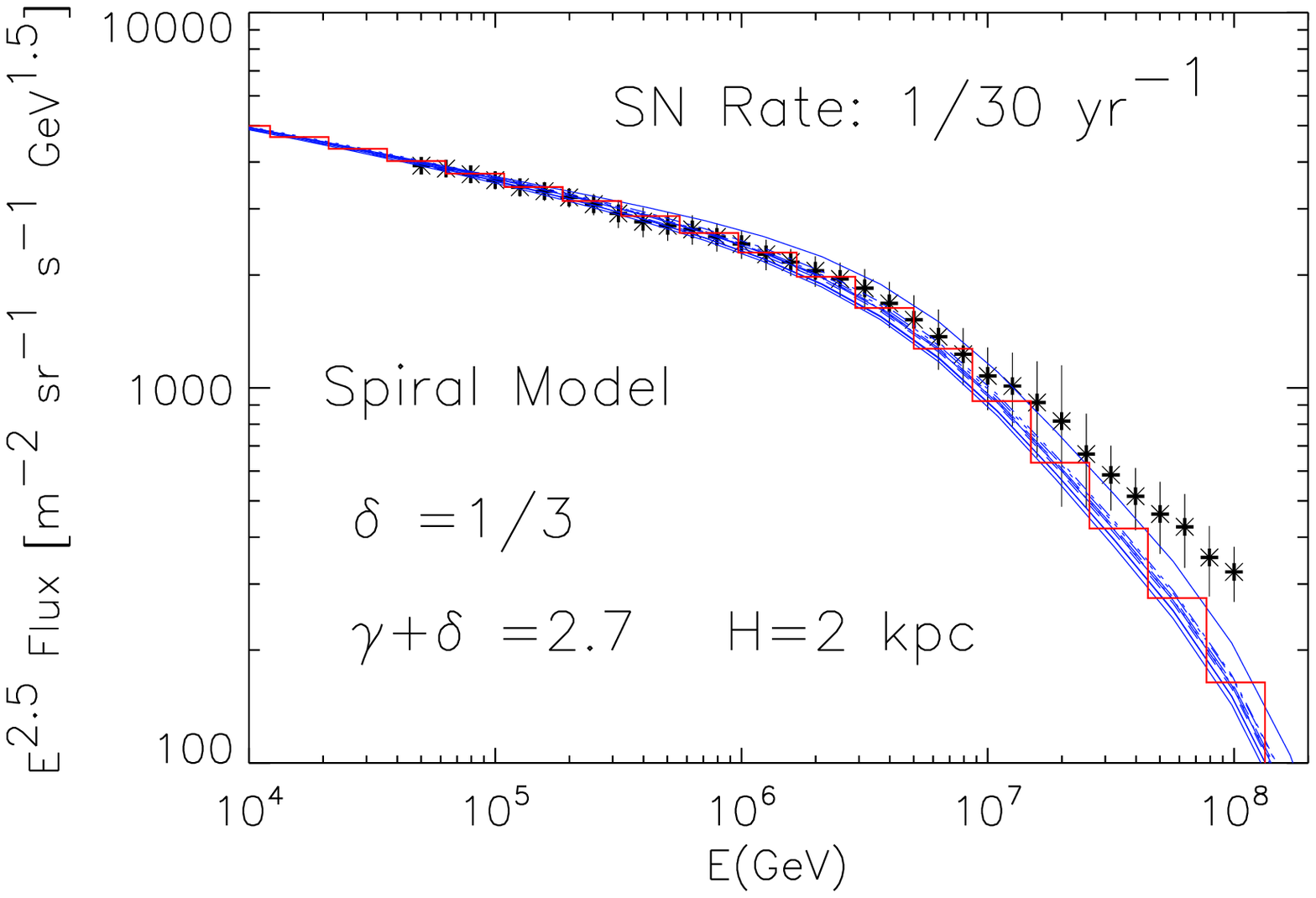}
\includegraphics[width=2.8in,angle=0]{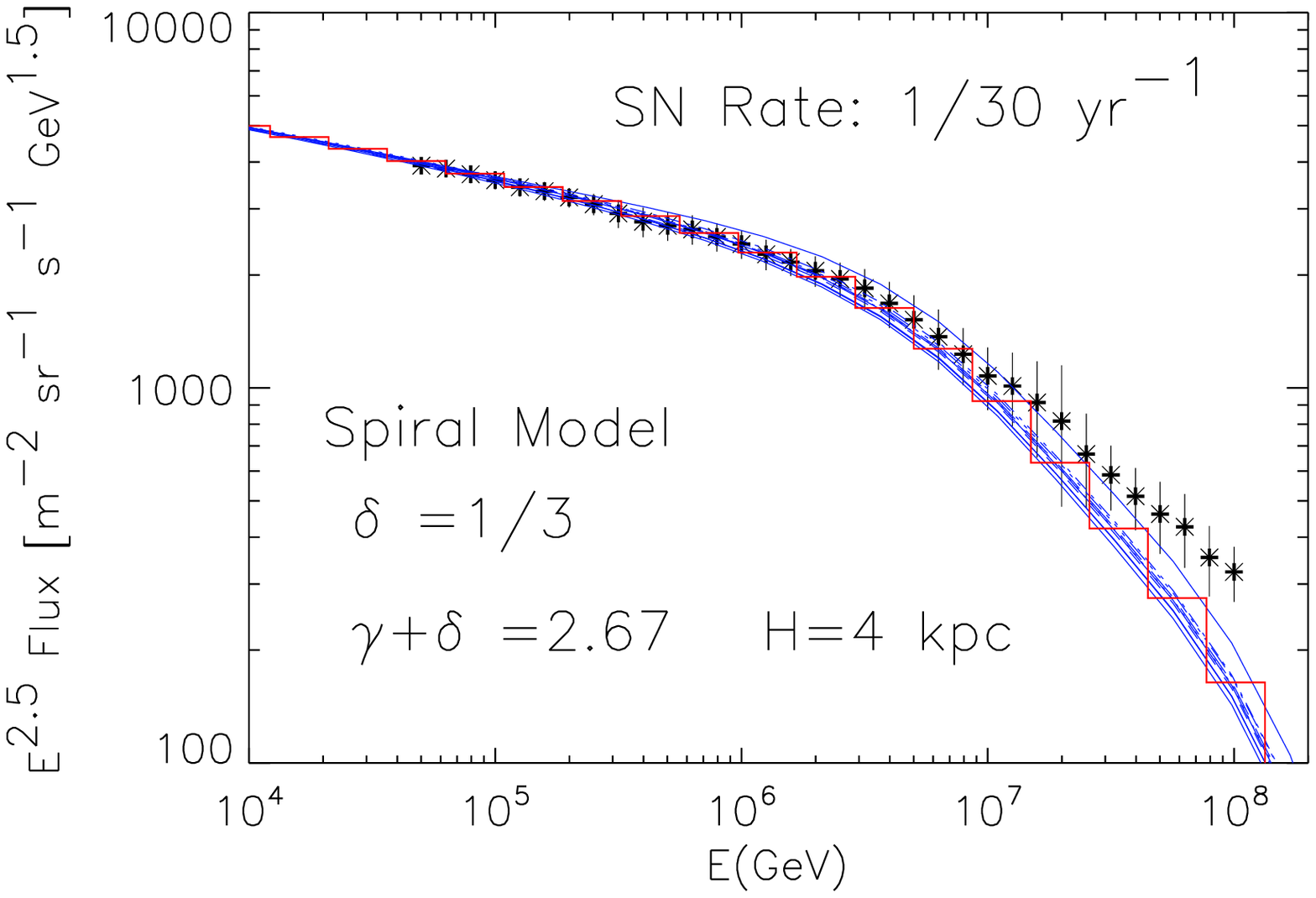}
\caption{Dependence of the all-particle spectrum on the realization of sources in the spiral model, for $H=2$ (left) and $H=4$ (right) kpc. In the left (right) panel we used $\gamma+\delta=2.7$ ($\gamma+\delta=2.67$).}  
\label{fig:H2vs4}
\end{figure}

This fact is illustrated in Fig.~\ref{fig:H2vs4} where we show the all-particle spectrum for $H=2$ (left) and $H=4$ kpc (right). For $H=4$ kpc, we have already seen that the spiral model reproduces the data with a spectral index at injection such that $\gamma_{obs}=2.67$: the same result that we obtained for the cylindrical model, independent of $H$. On the other hand, for  $H=2$ kpc$, \gamma_{obs}=2.67$ is only marginally compatible with the data, while $\gamma+\delta=2.7$ is found to be favorite. This is perfectly consistent with the fact that in the spiral model, for a given value of $\gamma+\delta$, heavy nuclei have a slightly harder spectrum. The difference in slope between the two cases, even if small, makes a sensible difference on the all-particle spectra because it sums up over many decades in energy.

In perfect analogy with the cylindrical case, the secular action of spallation leads also in this case to the flattening of the spectra of heavy elements for $\delta=1/3$. The He spectrum is again flatter than the protons' by the same amount as in the previous model, and again up to an energy of $\sim 10^5$ GeV. 

\subsection{Comments on the He hardening}

Two important observational findings recently appeared in the literature on cosmic rays: first, the sharp hardening found by ATIC-2 \cite{wefel08}, CREAM \cite{Ahn:2010p624} and PAMELA \citep{Adriani:2011p69} in both the proton and He spectra at $\sim 200$ GeV/nucleon; second, the fact that the He spectrum above this energy appears to be harder than the proton spectrum. The latter effect is observed at a very high level of significance by all three experiments mentioned above.

Taken at face value, the first finding alone represents a serious challenge to our attempts at achieving a comprehensive fit to all cosmic ray observables, an attempt that is routinely made by using GALPROP or similar propagation codes. Two possibilities come to mind: that the steep, low energy, part of the spectrum may be contributed by a so-far unknown nearby, probably old, cosmic ray source; or that at low energy cosmic rays experience a scattering rate that decreases with decreasing energy more than expected based on extrapolation from high energies, namely the diffusion coefficient has a stronger energy dependence with consequent steepening of the equilibrium spectra. 

The first scenario effectively decouples the B/C ratio from the measured spectral shape of primary cosmic rays that produce B through spallation reactions: the production of B depends on the mean Galactic CR spectrum, which in this case would be different, at low energies, from the spectrum at Earth. In this scenario it is virtually impossible to know the actual spectrum of CRs in the Galaxy, though some clues can be gathered from diffuse backgrounds, such as gamma rays and radio emission. It is worth stressing that in this case, propagation codes such as GALPROP cannot be easily used since they do not include the possibility of taking into account discrete sources.

In the second scenario one expects the dependence on energy of the grammage to be stronger at low energies than it is at high energies, thereby leading to a steepening of the equilibrium spectra of all cosmic ray components. The B/C ratio would reflect this complex situation, but would not provide any strong constraint on the grammage experienced by high energy particles. 

In both situations extrapolating functional forms from high to low energies appears unjustified and potentially dangerous in that it may lead to flawed conclusions. In a recent paper \cite{vladimirov2011}, among other things, the authors also comment on the results presented on a preliminary version of the present paper (which had been made publicly available as an arXiv preprint): the authors extrapolate our curves and conclusions to very low energies in order to compare predictions with B/C data where they are available. They conclude that if one interprets the He harder spectrum as a consequence of spallation, the grammage traversed by particles at lower energies leads to exceed the flux of antiprotons and the B/C ratio. However, antiproton measurements are only available at energies lower than 200 GeV \cite{Collaboration:2010p785}, namely outside the range of applicability of our model, which at those energies cannot even reproduce the CR spectrum, as already stressed above. On the other hand, for the B/C ratio there are two measurements, by CREAM \cite{Ahn:2008p1594} and TRACER \cite{TRACER2011} at energies higher than 1 TeV, but both have large error bars. If these two points are taken with the error bars shown in \cite{TRACER2011}, no inconsistency can be claimed between our model and the relevant data. The obvious conclusion is then the same that could be reached by observation of the spectral steepening suffered by all chemicals below $\sim$200 GeV/nucleon: the grammage cannot be extrapolated down in energy with the same slope that we assume at TeV energies. 

As far as the spallation-related explanation of the He hardening is concerned, we think that there is still enough uncertainty in the B/C ratio in the TeV range to warrant the benefit of doubt on whether this may be the correct explanation. In our view this scenario is certainly less {\it ad hoc} and more physically motivated than assuming artificial breaks in the source spectra or unknown source populations. To our knowledge this is the only attempt so far at understanding the new data on differential hardening in terms of well established propagation physics, rather than simply fitting them. 

Finally, a very important feature of this scenario is that it has the potential to be disproved not too far in future, through more accurate measurements of secondaries in the high energy range, with TRACER and AMS-2. While extrapolation to energy regions where the initial assumptions are known not to hold might lead to misleading conclusions, we think that any attempt should be made to test whether the spallation of He at TeV energies may lead to observables that are in contradiction with data in the same energy range.

\section{Additional scenarios}\label{sec:more}
Other physically relevant scenarios have been tested but did not induce particularly interesting effects on the spectrum and chemical composition: one is the continuous leakage of CRs from SNRs, as discussed in \S~\ref{sec:green}, and the other is the scenario in which supernova explosions occur dominantly in OB associations. Both these assumptions have negligible consequences in terms of changes induced in the spectrum and chemical composition of CRs observed at Earth. 

\section{The effect of extragalactic cosmic rays}\label{sec:extragal}

The spectrum and the chemical composition in the energy region above the knee may be substantially affected by a contribution from CRs of extragalactic origin. This is true in both descriptions of the transition from galactic to extragalactic that are presently considered as most likely, namely the electron-positron dip model \cite{Berezinsky:2005p1708} and the mixed composition scenario \cite{Allard:2005p1728}. Since the investigation of the transition is not among the purposes of the present paper, here, for simplicity, we illustrate this effect only for the case of the dip. We calculate the spectrum of extragalactic CRs as discussed in detail in \cite{Aloisio:2007p131} and normalize the absolute flux to the HiRes data (\cite{Sokolsky:2007p1741} and references therein). The injection spectrum of extragalactic CRs is assumed to be a power law with slope $-2.7$. No evolution of the source luminosity with redshift is taken into account since the only purpose of the calculation is to illustrate the effect, and not to explore the region of allowed parameters. As discussed in \cite{Aloisio:2007p131}, the dip model provides a suitable explanation of the observations provided that no more than $\sim 15\%$ of the extragalactic flux is contributed by He nuclei. Therefore we assume a pure proton composition of extragalactic CRs.

Even a small magnetic field in the intergalactic medium can suppress the flux of CRs coming from extragalactic sources, thereby introducing a low energy cutoff in the spectrum. Such low energy cutoff can also be intrinsic to the source if the sources harbor relativistic shocks: in this case the minimum energy of the accelerated particles is $\sim 4 \Gamma^{2} m_{p} c^{2}$, with $\Gamma$ the Lorentz factor of the shock. 

\begin{figure}[t]
\centering\leavevmode
\includegraphics[width=2.8in,angle=0]{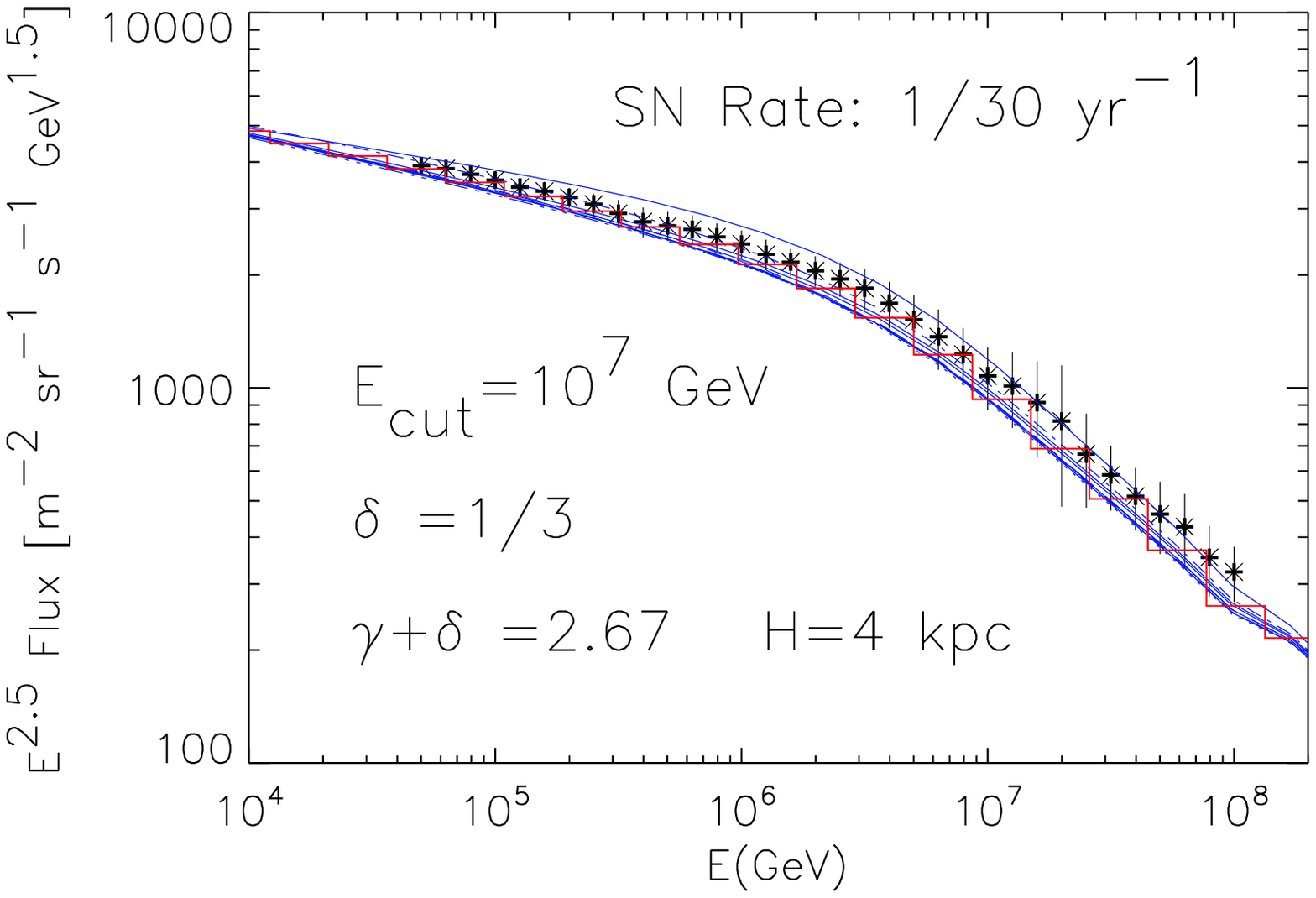}
\includegraphics[width=2.8in,angle=0]{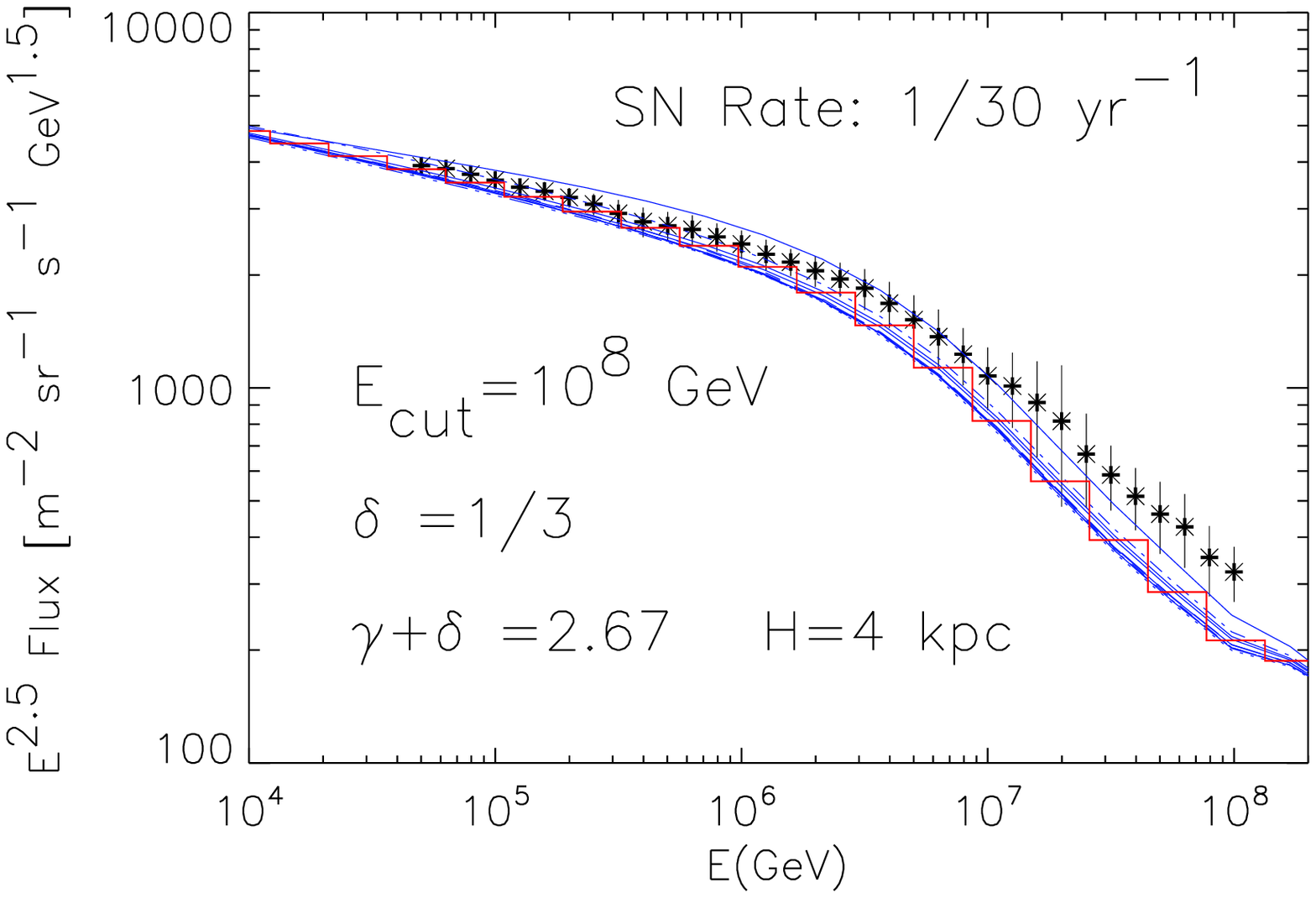}
\caption{All particle spectrum of CRs including the extragalactic component with $E_{cut}=10^{7}$ (left) and $E_{cut}=10^{8}$ GeV (right).}  
\label{fig:specEcut}
\end{figure}

\begin{figure}[t]
\centering\leavevmode
\includegraphics[width=2.8in,angle=0]{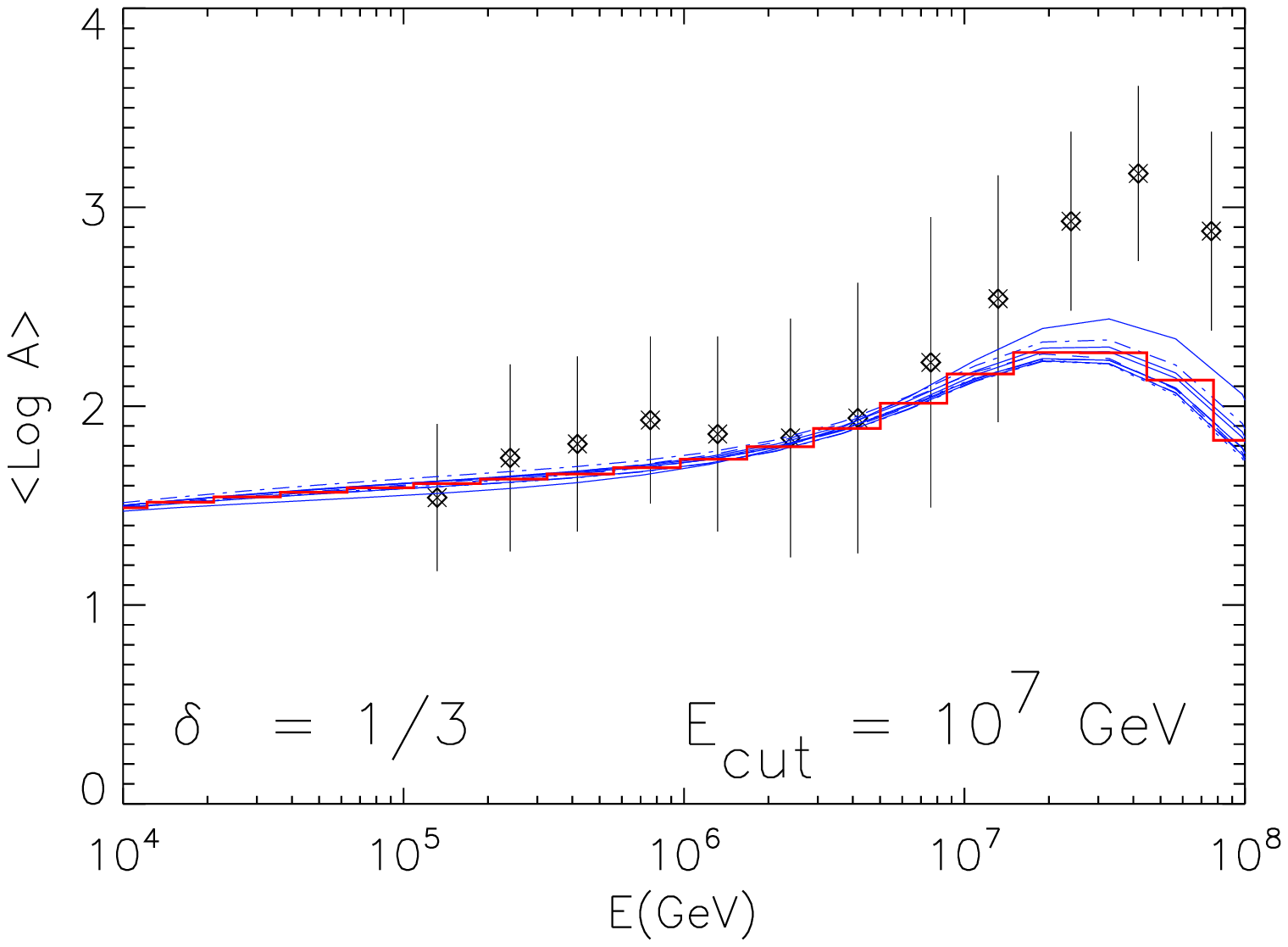}
\includegraphics[width=2.8in,angle=0]{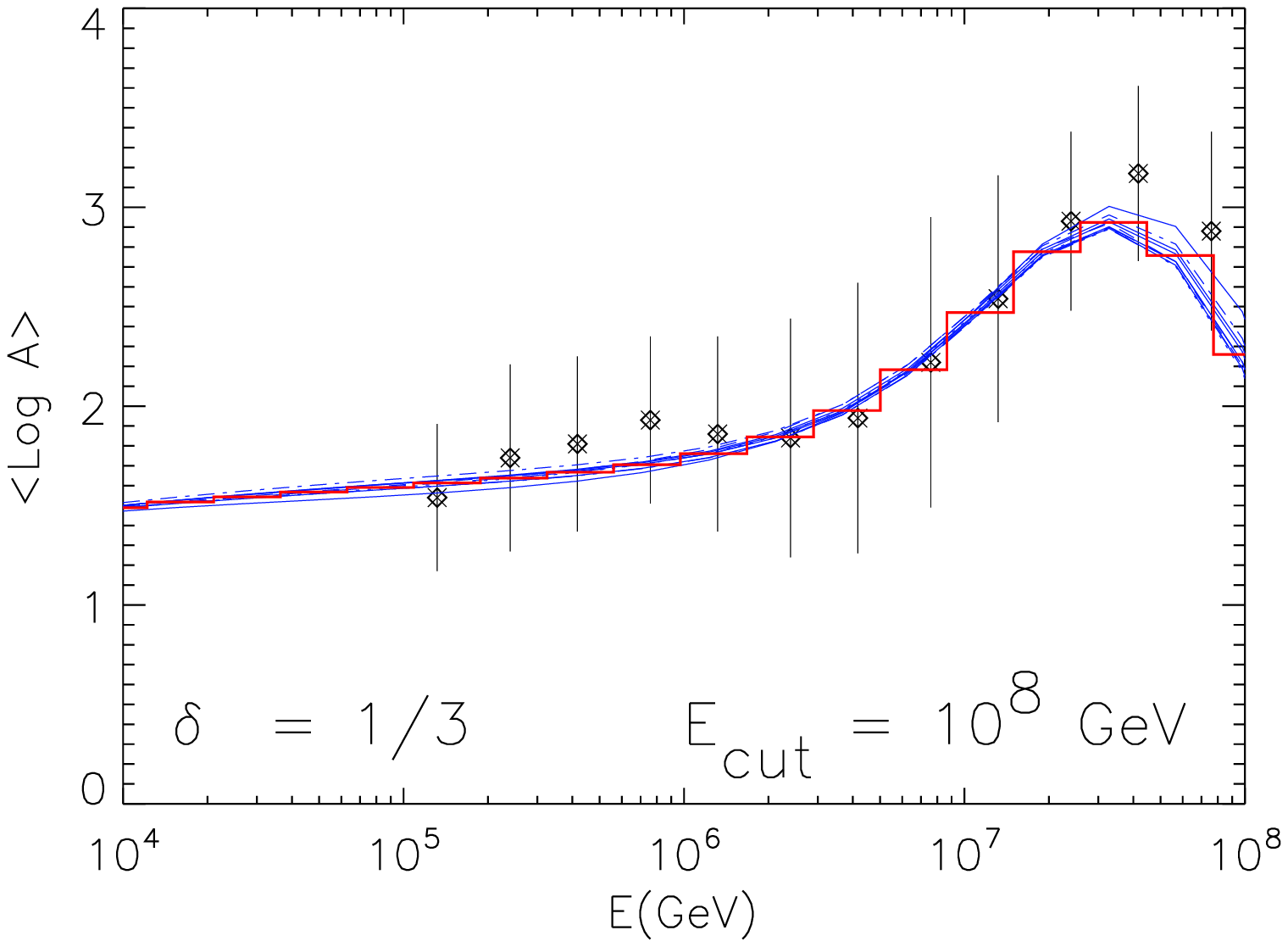}
\caption{Mean value of $\log A$ including the extragalactic component with $E_{cut}=10^{7}$ (left) and $E_{cut}=10^{8}$ GeV (right).}  
\label{fig:logAEcut}
\end{figure}

We artificially model this low energy suppression with an exponential cutoff at $E_{cut}=10^{7}$ GeV and $10^{8}$ GeV. The all-particle spectrum for these two values of $E_{cut}$ is plotted in Fig.~\ref{fig:specEcut}: one can see that the superposition of the Galactic and extragalactic CR spectra allows us to provide a smooth fit to the available data in both cases. In Fig.~\ref{fig:logAEcut} we plot the mean logarithmic mass, $\langle \log A \rangle$, as a function of energy for the two values of $E_{cut}$: the goodness of the fit is negatively affected by an extragalactic proton spectrum extending too low in energy, therefore if we are to believe this type of interpretation of the transition, and take strictly the data on the mean log mass, then the low energy cutoff of the extragalactic CR spectrum has to be at $\gtrsim 10^{8}$ GeV. At very high energy, Fig.~\ref{fig:logAEcut} suggests that the transition has to occur through mixing of different chemicals. However one has to keep in mind that the measurements of the composition in the transition region are still affected by large systemic errors which do not allow very firm conclusions in this important energy range, as also discussed in \cite{Aloisio:2008p126}. Recent data from HiRes \cite{Sokolsky:2007p1741} and Telescope Array \cite{Thomson:2010p1545} seem to find a chemical composition at $\sim 10^{18}$ eV dominated by protons, which is not reflected in the data used in Fig.~\ref{fig:logAEcut}. The most recent data on the chemical composition as observed with the Pierre Auger Observatory also show a proton dominated composition at $\sim 10^{18}$ eV, but gradually shifting to heavier composition at higher energies (see for instance \cite{Roulet:2011p1892}).

Interestingly enough our results do not require an additional component (possibly super-heavy Galactic nuclei invoked by \cite{Hillas:2005p171} and \cite{Horandel:2004p1543}, see also \cite{deDonato:2009p1547}) to fit the shape of the all-particle spectrum. The need for such an extra component was probably arising from assuming that the maximum energy of protons accelerated in SNRs equal the energy where the knee appears. In our calculations, as discussed above, the maximum energy of protons that best fits the all particle spectrum is $\sim 6\times 10^{6}$ GeV, well above the knee, but the knee itself appears to be mainly contributed by Helium. The shape of the cutoff in the spectra of individual chemicals also affects the need for extra-components. 

\section{Summary and Conclusions}\label{sec:conclusion}
In this paper we have tried to make progress in the quest for the origin of CRs, trying to clarify the requirements on propagation and source spectra that come from observations of CRs at Earth. 

We computed the spectrum and chemical composition of CRs at Earth assuming that SNRs are the primary sources of galactic CRs and taking into account the discrete distribution in both space and time of these sources. Each source is assumed to inject protons and nuclei with a power law spectrum up to some maximum energy, above which the spectrum is exponentially cut off. The power law index and maximum energy are assumed to be the same for all sources and are determined from observations, with $E_{max}$ simply scaling with rigidity for nuclei heavier than H. The propagation of CRs from their sources to Earth is treated through a Green function approach allowing for a completely general dependence on space and time of the injected spectrum, as well as taking into account spallation. The diffusion coefficient has been taken constant in space within the galactic halo and has a power-law dependence on rigidity alone with index $\delta=1/3$ or $\delta=0.6$. In terms of its normalization, the diffusion coefficient is assumed to scale with the size of the halo, following the findings of other propagation codes, such as GALPROP or DRAGON.

We have considered a few different scenarios for the distribution of sources. As far as their spatial distribution is concerned, we have carried out the calculations both for a smooth distribution of SNRs within the galactic disk (cylindrical model) and for the case in which SNRs follow the spiral structure of the Galaxy (spiral model). We have also considered the possibility that SNRs are clustered in superbubbles but we find that in terms of spectrum and chemical composition this assumption makes negligible difference.

Concerning the dependence on time of the injected spectrum, we have considered two different scenarios: a case in which all the CRs a SNR has ever accelerated are released at the end of its evolution, and a case in which, at all times through the Sedov phase of the SNR evolution, there is continuous leakage of CRs at the maximum energy from upstream of the decelerating blast wave, with the maximum energy decreasing with time down to $\sim 1GeV$. These two different models for the time dependence of the injection lead to very similar outcomes in terms of spectrum and chemical composition, once the spatial distribution of the sources and the energy dependence of the diffusion coefficient are fixed.

Our main findings in the present paper are summarized in the following.
First of all the ability of our approach at taking into account the discreteness of the source distribution allows us to illustrate that the all-particle CR spectrum at Earth is not in general guaranteed to be a power law. In particular if the diffusion coefficient depends on rigidity as $E^\delta$ with $\delta=0.6$ a number of wiggles are likely to appear in the spectrum, due to the contribution of nearby sources. This fact also suggests that this value of $\delta$ will lead to large levels of anisotropy. 

For a diffusion coefficient scaling with $E^{1/3}$, the CR spectrum and chemical composition observed at Earth can be reproduced in a satisfactory way in both the cylindrical and spiral model of source distribution in the Galaxy. The required maximum energy of the accelerated particles is $E_{max}=6\times 10^6$ GeV for protons, reasonably close to (but somewhat higher than) what can be expected from a SNR based on the non linear theory of diffusive shock acceleration. The knee in the all-particle spectrum results from the scaling of the maximum particle energy with rigidity. The required power-law index of the injected particles is the same in both scenarios for a halo size $H=4\ kpc$, namely $\gamma=2.34$. A slightly steeper power-law is required instead in the spiral case if the magnetized halo has size $H=2\ kpc$. In this latter case one finds $\gamma=2.37$, and the small difference is induced by the need of compensating for the effects of spallation that lead to a hardening of the spectrum at low energy, with a deficiency of low energy particles that in this scenario cannot be supplied by nearby sources due to the inter-arm position of the solar system.

These spectral indices represent at present the most problematic aspect of the supernova remnant paradigm for the origin of CRs. The required acceleration efficiency is relatively low, for $D\propto E^{1/3}$ and $\cR=1/30\ yr^{-1}$ being only of order $6-7\%$.

Proper account of the effects of spallation leads to a hardening of the spectrum of heavy elements at low energy, even when this process is not dominant over escape. This is the key element that leads us to expect that, for $\delta=1/3$, He nuclei have a flatter spectrum than protons at low energies and dominate the all-particle spectrum at the knee, in qualitative agreement with the recent findings of CREAM \cite{Ahn:2010p624,Ahn:2009p1595} and PAMELA \cite{Adriani:2011p69}. The slopes of the protons and Helium spectra that we find are $\sim 2.67$ and $\sim 2.6$ respectively, to be compared with the measured values of $2.66\pm 0.02$ and $2.58\pm 0.02$ \cite{Ahn:2010p624}. Our calculations show that for $\delta=0.6$ this effect disappears, that might indeed suggest that Galactic diffusion is characterized by a low value of $\delta$.

The prediction of the hardening in the spectra of nuclei compared with that of protons is one of the most important results of our calculations. As noted by \cite{vladimirov2011}, if extrapolated down to much lower energies this finding may cause excessive B and antiprotons production. However, a more straightforward consequence of such a naive extrapolation would be that the spectra of H and He would not reflect the observed steepening of both components at rigidities $<200$ GeV/nucleon, which we make no attempt at explaining. In other words, what this is suggesting is simply that at low energies something new is happening. Currently this is not yet explained in physical terms: the data can only be reproduced by introducing artificial breaks in either the injection spectra or the diffusion coefficient, or both. We limited ourselves to consider the high energy range and the simplest possible physical scenario for the very reason that we did not want to introduce additional parameters before having an idea of the nature of the underlying phenomena. After all, if one applied the same approach to the knee in the CR spectrum, assuming breaks in the injection spectra and/or in the diffusion coefficient could certainly lead to a good fit to the data, but would make little contribution to our understanding of the origin of this spectral feature.

Currently we cannot explain the fact that the spectra of all species flatten rather abruptly at $\sim 200$ GeV/n as found by the PAMELA experiment \cite{Adriani:2011p69}. However we do not share the opinion of the authors claiming the failure of the SNR paradigm or the existence of an unknown accelerator. We rather think that the SNR paradigm is in fact more complex than usually assumed in doing these claims, and that its consequences are not yet so well understood as sometimes people would like to believe.

At high energies, above the knee, we underpredict the all-particle spectrum and slightly overpredict the mass mean Log of CRs. This seems to suggest that a lighter CR component is probably contributed at those energies by extragalactic sources. Since fitting the details of the observed CR spectrum at high energy was not our main purpose in this paper, we modeled this extragalactic component only within the scenario of the Dip \cite{Berezinsky:2005p1708}, in which the extragalactic CR composition is dominated by protons. In this case we find that the spectrum can be fitted very well, while the chemical composition suggests that the extragalactic component should have a low energy cutoff at energies larger than $10^8$ GeV. Fitting the all-particle spectrum does not require any additional heavy component as previously postulated in Refs. \cite{Hillas:2005p171} and \cite{Horandel:2004p1543}.

Our general conclusion is that if the diffusion coefficient in the Galaxy scales with energy as $E^{1/3}$ SNRs can account for the all-particle spectrum and chemical composition of CRs detected at Earth, provided that they are able to accelerate protons up to $\sim 6 \times 10^6$ GeV and together give spectra as steep as $E^{-\gamma}$ with $\gamma=2.3-2.4$. This is indeed a bit of a challenge for the non-linear theory of diffusive shock acceleration, which is believed to describe the acceleration process in this context. On the one hand, efficient acceleration leads to hard (and even concave) particle spectra, at odds with the relatively large values of $\gamma$ obtained above. On the other hand, the inferred maximum energy can only be achieved if the magnetic field is largely amplified with respect to its average value in the ISM. Efficient amplification is usually associated with efficient particle acceleration, whereby the difficulties at reconciling steep spectra and high maximum energies. These difficulties might be overcome in the context of NLDSA if the velocity of the scattering centers that enters the transport equation is that computed in the amplified magnetic field, rather than the unperturbed Alfv\`en velocity \cite{Caprioli:2011p1899}. In fact, the use of the unperturbed velocity derives from a strict application of quasi-linear theory, which is however questionable in the presence of such large levels of field amplification as those usually found appropriate for young SNRs. In addition, as soon as the Alfv\`en velocity in the self-generated magnetic field is considered, the acceleration efficiency is much reduced, while it may still happen that the maximum proton energy reaches the knee. This recent result is very interesting, especially in the light of our findings in this paper, that the required acceleration efficiency is only about $5\%$ in our best fit scenario, suggesting that the acceleration process has to guarantee decoupling between a large maximum energy and efficient particle acceleration. 

Another possibility to reconcile steep spectra of the accelerated particles and a high maximum energy might be related to the shock obliquity. The theory of NLDSA has so far been developed only for parallel shocks, while particle acceleration at perpendicular and more generally oblique shocks has not been devoted the same amount of attention. Perpendicular shocks are fast accelerators, and in principle allow to reach higher energies than parallel shocks. This is especially interesting in that the value we require for the maximum energy of protons is close to the upper limit of what can be achieved within NLDSA at parallel shocks. Whether the required combination of steep spectra and high maximum energies can be achieved at highly oblique shocks is an issue that deserves further investigation. 

\bibliographystyle{JHEP}
\bibliography{crbib}

\end{document}